\documentclass[preprint,12pt]{elsarticle}

\usepackage{amssymb}
\usepackage{amsmath}
\usepackage{orcidlink}
\usepackage{booktabs}
\usepackage{multirow}
\usepackage{multicol}
\usepackage[capitalize]{cleveref}
\usepackage{pgfplots}
\usepackage{float}
\usepackage{tikz}
\renewcommand{\d}{\,\textnormal{d}}
\newtheorem{remark}{Remark}

\usepackage{changepage}
\usepackage{geometry}
\usepackage{pgfplots}

\usepackage{multirow}
\newcommand*\ttvar[1]{\texttt{\expandafter\dottvar\detokenize{#1}\relax}}
\newcommand*\dottvar[1]{\ifx\relax#1\else
	\expandafter\ifx\string_#1\string_\allowbreak\else#1\fi
	\expandafter\dottvar\fi}

\newcommand{\out}[1]{}

\usepackage{todonotes}
\newboolean{SHOWCOMMENTS}
\setboolean{SHOWCOMMENTS}{true}

\ifthenelse{\boolean{SHOWCOMMENTS}}{
	\newcommand{\tdMK}[1]
	{\todo[color=blue!25,inline]{\footnotesize{\bf Martin:} #1}}
}{
	\newcommand{\tdMK}[1]{}
}
\newcommand{\MK}[1]{{\color{blue}#1}}
\ifthenelse{\boolean{SHOWCOMMENTS}}{
	\newcommand{\tdJL}[1]
	{\todo[color=blue!25,inline]{\footnotesize{\bf Julian:} #1}}
}{
	\newcommand{\tdJL}[1]{}
}
\ifthenelse{\boolean{SHOWCOMMENTS}}{
	\newcommand{\tdCK}[1]
	{\todo[color=red!25,inline]{\footnotesize{\bf Carola:} #1}}
}{
	\newcommand{\tdCK}[1]{}
}
\ifthenelse{\boolean{SHOWCOMMENTS}}{
	\newcommand{\tdPL}[1]
	{\todo[color=teal!40!white,inline]{\footnotesize{\bf Philippe:} #1}}
}{
	\newcommand{\tdPL}[1]{}
}

\journal{Elsevier}

\begin{document}
	
	\begin{frontmatter}
		
		
		\title{Memory- and compute-optimized geometric multigrid \textit{GMGPolar} for curvilinear coordinate representations -- Applications to fusion plasma}

		\author[a,label1]{Julian Litz}
		\fntext[label1]{Current address: Institute of Climate and Energy Systems — Energy Systems Engineering (ICE-1), Forschungszentrum Jülich GmbH, 52425 Jülich, Germany}
		\author[b]{Philippe Leleux}
		\author[c]{Carola Kruse}
		\author[d]{Joscha Gedicke}
		\author[a,e]{Martin~J.~K\"uhn\orcidlink{0000-0002-0906-6984}}
		\ead{martin.kuehn@dlr.de}
		
		\address[a]{{German Aerospace Center (DLR), Institute of Software Technology, Department for High-Performance Computing}, 
			{Linder H\"ohe}, 
			{Cologne}, 
			{51147}, 
			{Germany}, 
		}
		\address[b]{{Laboratoire d'Analyse et d'architecture des Systèmes (LAAS), équipe TSF}, 
			{7 avenue du Colonel Roche BP 54200}, 
			{Toulouse cedex 4}, 
			{31031}, 
			{France}, 
		}
		\address[c]{{Parallel Algorithms Team, CERFACS (Centre Europ\'een de Recherche et de Formation Avanc\'ee en Calcul Scientifique)}, 
			{42 Avenue Gaspard Coriolis}, 
			{Toulouse Cedex 01}, 
			{31057}, 
			{France}, 
			
		}
		\address[d]{ {Institut f\"ur Numerische Simulation, Universit\"at Bonn}, 
			{Friedrich-Hirzebruch-Allee 7}, 
			{Bonn}, 
			{53115}, 
			{Germany}. 
		}
		
		\address[e]{{Life \& Medical Sciences Institute (LIMES) and Bonn Center for Mathematical Life Sciences, University of Bonn}, 
			{Bonn}, 
			{53115}, 
			{Germany}, 
		}
		
		\begin{abstract}
			Tokamak fusion reactors are actively studied as a means of realizing energy production from plasma fusion. However, due to the substantial cost and time required to construct fusion reactors and run physical experiments, numerical experiments are indispensable for understanding plasma physics inside tokamaks, supporting the design and engineering phase, and optimizing future reactor designs. 
			Geometric multigrid methods are optimal solvers for many problems that arise from the discretization of partial differential equations. It has been shown that the multigrid solver GMGPolar solves the 2D gyrokinetic Poisson equation in linear complexity and with only small memory requirements compared to other state-of-the-art solvers.
			In this paper, we present a completely refactored and object-oriented version of GMGPolar which offers two different matrix-free implementations. Among other things, we leverage the Sherman-Morrison formula to solve cyclic tridiagonal systems from circular line solvers without additional fill-in and we apply reordering to optimize cache access of circular and radial smoothing operations. With the \textit{Give} approach, memory requirements are further reduced and speedups of four to seven are obtained for usual test cases. For the \textit{Take} approach, speedups of 16 to 18 can be attained. \MK{In an additionally experimental setup of using GMGPolar as a preconditioner for conjugate gradients, this speedup could even be increased to factors between 25 and 37.}
		\end{abstract}

		\begin{highlights}
			
			\item Realizing a very fast, memory-minimized and higher order solver for the gyrokinetic Poisson equation on 2D cross sections of tokamak geometries
			\item A fully refactored, object‑oriented and user-friendly geometric multigrid framework for curvilinear coordinate representations
			\item Cache-access optimized multigrid line smoothers for curvilinear coordinates
			\item Achieved $16\times\,-\,18\times$ speedups and a one-third memory reduction compared to the prior state-of-the-art version
			\item \MK{Experimental speedup factors of 25--37 when used as Krylov preconditioner}
			
		\end{highlights}

		
	\end{frontmatter}
	

		\section{Introduction}
		
		Tokamak fusion reactors are one of the most promising approaches for realizing energy production from plasma fusion. 
		However, due to substantial cost and time to construct fusion reactors and run physical experiments, numerical experiments are indispensable to understand plasma physics inside tokamaks, to support and speed up the engineering phase, and to optimize future reactor designs. 
		
		To model and simulate the particular physics, the gyrokinetic framework is used by many authors; see, e.g., \cite{GENE-3D,COGENT2020,EUTERPE,ORB5,grandgirard20165d,bouzat2018targeting}. The corresponding five-dimensional problem to be solved contains three dimensions for the torus geometry and two dimensions for the velocity~\cite{grandgirard20165d}.
		At each time step, the simulation requires solving a 5D Vlasov equation for each species, along with a 3D Poisson-like equation that enforces quasi-neutrality.
		While of smaller dimension, computing the solution of this three-dimensional system can deteriorate the overall performance and scaling of simulation codes~\cite{grandgirard20165d}. 
		While some solvers, such as GENE-3D~\cite{GENE-3D},~{COGENT~\cite{COGENT2020}}, or EUTERPE~\cite{EUTERPE}, solve this system directly, other codes such as GYSELA~\cite{grandgirard20165d}, and ORB5~\cite{ORB5} solve a large number of two-dimensional equations on cross sections of the tokamak; cf.~\cref{fig:tokamak_cross_270} for a visualization of a tokamak geometry with several cross sections. 
		The type and form of the 2D cross sections differ for various publications (see also the next section).
		As the formulation by curvilinear coordinates poses an additional difficulty at the section of the separatrix, recently, in~\cite{vidal2025local}, a multi-patch geometry for a decomposed domain with an X-point was presented.
		
		In~\cite{KKR22}, the taylored geometric multigrid method~\textit{GMGPolar} was proposed to efficiently solve the resulting subproblems on hundreds or thousands of cross sections, repeated over thousands or millions of time steps. From the beginning, GMGPolar has been designed to be scalable, allow higher order approximations, and reduce the memory footprint to a minimum.
		In~\cite{LSK25}, a matrix-free C++ implementation using shared memory parallelism through OpenMP was presented. 
		It was furthermore shown that the GMGPolar algorithm is optimal in the sense that it has linear asymptotic complexity, i.e., the number of floating point operations (FLOP) for the solution process is a linear function of the number of degrees of freedom. In~\cite{BLK23}, GMGPolar was compared to other state-of-the-art solvers for the gyrokinetic Poisson-like equation on tokamak cross sections. 
		GMGPolar was found to have the smallest memory requirements and to offer a compromise between relatively fast execution and high order of approximation. 
		
		In this article, we present a completely refactored version 2 of GMGPolar~\cite{gmgpolar_code}. 
		The novel version aligns the multigrid components with cache lines, optimizes the compromise between the storage and recomputation of (costly) function evaluations, underwent low level performance engineering through, e.g., function inlining, and boosts the parallel scalability through a substantial reduction of synchronization and waiting times. 
		In addition, GMGPolar now implements $F$-cycles and full multigrid to speed up the convergence.
		
		The article is structured as follows. In~\cref{sec:model_and_gmg}, we present the considered model problem together with the mathematical background and multigrid components of GMGPolar. 
		In~\cref{sec:oore}, we present the object-oriented design of the geometric multigrid algorithm. 
		Here, we focus particularly on optimizing memory usage and cache accesses, improving parallel scalability, and on new multigrid features such as full multigrid. 
		We provide extensive numerical results in~\cref{sec:num} before concluding with~\cref{sec:conc}.

		\section{Model problem and principles of GMGPolar}\label{sec:model_and_gmg}
		
		\subsection{Model problem and geometry}
		
		With some simplifications, as explained in~\cite{BLK23}, we consider the following model problem for a 2D Poisson-like equation
		\begin{align}\label{eq:poisson}
			\begin{alignedat}{3}
				-\nabla \cdot (\alpha \nabla u) + \beta u &= f     &&\quad \text{in}\quad &&\Omega,\\
				u &= u_D   &&\quad \text{on} \quad &&\partial\Omega,
			\end{alignedat}
		\end{align}
		which arises in the description of quasi-neutrality. 
		Here, $\Omega \subset \mathbb{R}^2$ is a disk-like domain, and $f:~\Omega \rightarrow \mathbb{R}$, with $f \in \mathcal{C}^0(\overline{\Omega})$, is the right-hand side. 
		The functions $\alpha, \beta:~\Omega \rightarrow \mathbb{R}$ are coefficients corresponding to \emph{density profiles}, with $\alpha \in \mathcal{C}^1(\Omega) \cap \mathcal{C}^0(\overline{\Omega})$ and $\beta \in \mathcal{C}^0(\overline{\Omega})$. 
		Furthermore, we prescribe Dirichlet boundary conditions with $u_D \in \mathcal{C}^0(\partial\Omega)$.
		
		\begin{figure}[h]
			\centering
			\includegraphics[width=0.85\linewidth,clip,trim={20cm 12cm 20cm 15cm}]{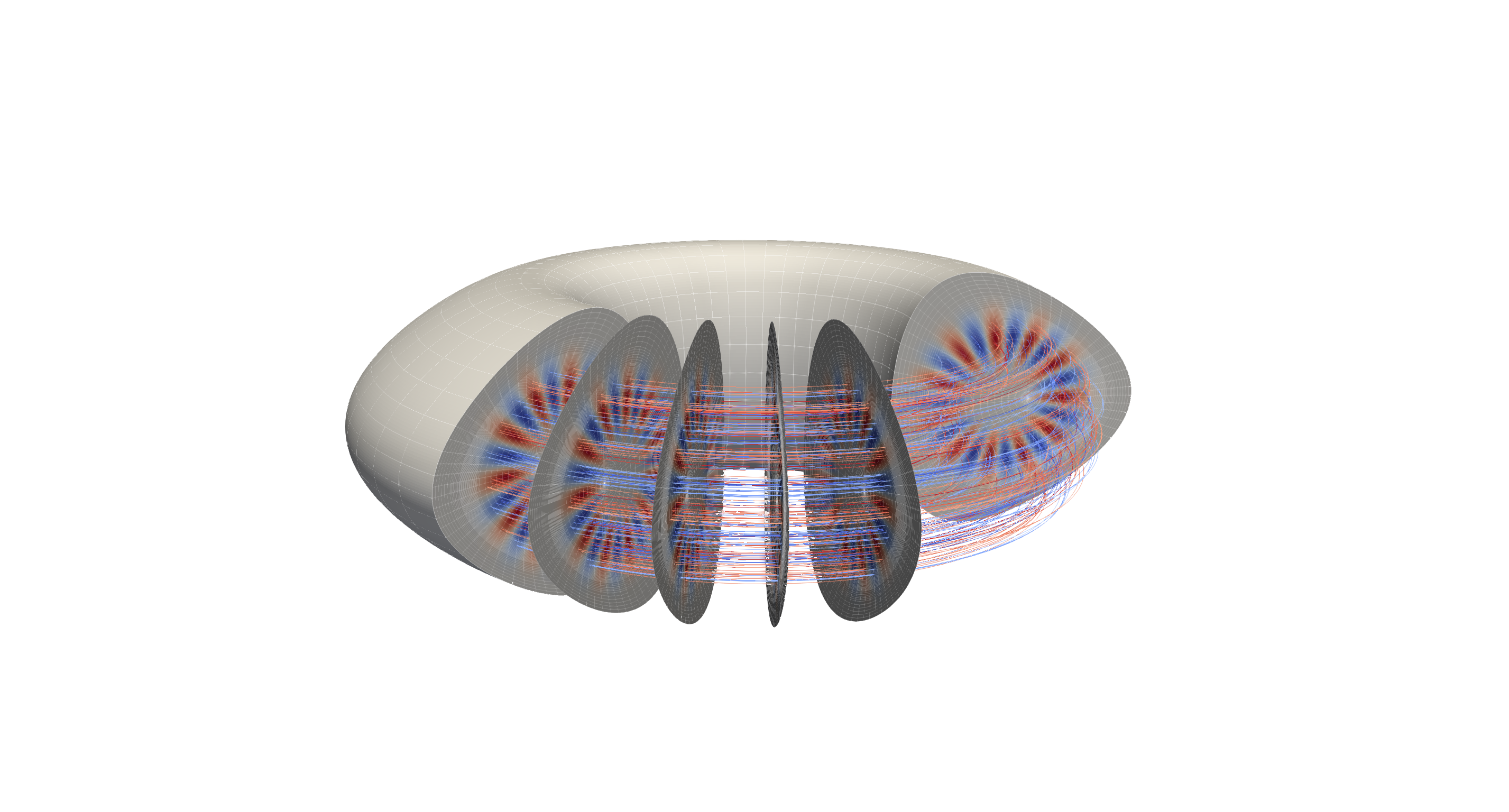}
			\caption{\textbf{Particular solution of~\eqref{eq:poisson} on a 3D tokamak with Culham geometry.} 
				The figure shows a 3D tokamak geometry spanning over 220 degrees in toroidal direction and clipped over the remaining 140 degrees. It furthermore show several 2D cross sections in  the clipped part of the geometry. 
			}
			\label{fig:tokamak_cross_270}
		\end{figure}
		
		We next introduce the energy functional
		\begin{align}\label{eq:functional}
			J(u):=\int_{\Omega}\frac{1}{2}\alpha|\nabla u|^2+\frac{1}{2}\beta u^2-fu \, \d(x,y),
		\end{align}
		from which we obtain a symmetric linear system after a finite difference discretization as described further below.
		We note that the weak formulation of problem~\eqref{eq:poisson} is equivalent to the minimization of the energy functional $J(u)$ in~\eqref{eq:functional} over a suitable Sobolev space, prescribing the boundary conditions $u_D$.
		
		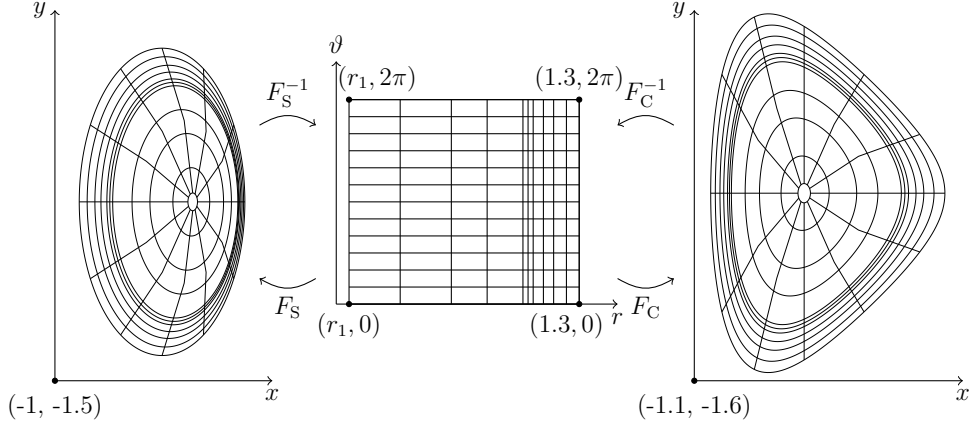
\begin{figure}
			\begin{tikzpicture}[scale=0.34,every node/.style={scale=0.8}]
				
				\begin{axis}[
					axis equal,
					xmin=-1, xmax=1,
					ymin=-2, ymax=2,
					hide axis,
					ticks=none,
					at={(-300,-200)},
					scale=2.5
					]
					\foreach \phi in {0,30,...,330} {
						\addplot [
						domain=0.075:1.3,
						samples=100,
						thin
						]
						(
						{0 + (1 - 0.3)*x*cos(\phi) - 0.2*x^2},
						{0 + (1 + 0.3)*x*sin(\phi)}
						);
					}
					
					\foreach \r in {0.075, 0.28888889, 0.57777778, 0.78, 0.98222222, 1.01111111, 1.04, 1.09777778, 1.15555556, 1.22777778, 1.3} {
						\addplot [
						domain=0:360,
						samples=200,
						thin
						]
						(
						{0 + (1 - 0.3)*\r*cos(x) - 0.2*\r^2},
						{0 + (1 + 0.3)*\r*sin(x)}
						);
					}
				\end{axis}
				
				\draw[->] (-7.5,-7) -- (1,-7) node[below] {$x$};
				\draw[->] (-7.5,-7) -- (-7.5,7.5) node[left] {$y$};
				
				\draw[fill](-7.5,-7) circle(0.1cm);
				\node at(-7.5,-8) {(-1, -1.5)};

				\draw[->] (0.5,3) .. controls +(30:1cm) and +(150:1cm) .. (2.7,3) node[above, midway, sloped] {$F_{\text{S}}^{-1}$};
				\draw[<-] (0.5,-3) .. controls +(-30:1cm) and +(210:1cm) .. (2.7,-3) node[below, midway, sloped] {$F_{\text{S}}$};

				\begin{axis}[
					axis equal,
					xmin=-1.6, xmax=0.5,
					ymin=-2.1, ymax=2.1,
					hide axis,
					ticks=none,
					at={(350,-200)},
					scale=2.5
					]
					
					\pgfmathsetmacro{\eps}{0.3}
					\pgfmathsetmacro{\invEps}{1/\eps} 
					\pgfmathsetmacro{\ecc}{1.4}
					\pgfmathsetmacro{\xi}{1.0114}
					
					\foreach \phi in {0,30,...,330} {
						\addplot [
						domain=0.075:1.3,
						samples=100,
						thin
						]
						(
						{ (1 / 0.3) * (1 - sqrt(1 + 0.3 * (0.3 + 2 * x * cos(\phi)))) },
						{ (1.4 * 1.0114 * x * sin(\phi)) / 
							(1 + 0.3 * (1 / 0.3) * (1 - sqrt(1 + 0.3 * (0.3 + 2 * x * cos(\phi))))) }
						);
					}
					
					\foreach \r in {0.075, 0.28888889, 0.57777778, 0.78, 0.98222222, 1.01111111, 1.04, 1.09777778, 1.15555556, 1.22777778,1.3} {
						\addplot [
						domain=0:360,
						samples=200,
						thin
						]
						(
						{ (1 / 0.3) * (1 - sqrt(1 + 0.3 * (0.3 + 2 * \r * cos(x)))) },
						{ (1.4 * 1.0114 * \r * sin(x)) / 
							(1 + 0.3 * (1 / 0.3) * (1 - sqrt(1 + 0.3 * (0.3 + 2 * \r * cos(x))))) }
						);
					}
				\end{axis}
				
				\draw[->] (17.5,-7) -- (28,-7) node[below] {$x$};
				\draw[->] (17.5,-7) -- (17.5,7.5) node[left] {$y$};
				
				\draw[fill](17.5,-7) circle(0.1cm);
				\node at(17.5,-8) {(-1.1, -1.6)};

				\draw[<-] (14.5,3) .. controls +(30:1cm) and +(150:1cm) .. (16.7,3) node[above, midway, sloped] {$F_{\text{C}}^{-1}$};
				\draw[->] (14.5,-3) .. controls +(-30:1cm) and +(210:1cm) .. (16.7,-3) node[below, midway, sloped] {$F_{\text{C}}$};

				\draw[->] (3.5, -4) -- (14.5, -4) node[below] {$r$};
				\draw[->] (3.5, -4) -- (3.5, 5.5) node[above] {$\vartheta$};
				
				\draw(4,-4) -- (13,-4) -- (13,4) -- (4,4) -- (4,-4);
				\draw(6,-4) -- (6,4);
				\draw(8,-4) -- (8,4);
				\draw(9.4,-4) -- (9.4,4);
				\draw(10.8,-4) -- (10.8,4);
				\draw(11,-4) -- (11,4);
				\draw(11.2,-4) -- (11.2,4);
				\draw(11.6,-4) -- (11.6,4);
				\draw(12,-4) -- (12,4);
				\draw(12.5,-4) -- (12.5,4);
				\draw(13,-4) -- (13,4);
				\foreach \i in {0,...,12} {
					\pgfmathsetmacro{\y}{-4 + \i * (4 - (-4)) / 12}
					\draw (4, \y) -- (13, \y);
				}
				
				\node[below] at (4,-4) {$(r_1,0)$};
				\draw[fill](4,-4) circle(0.1cm);
				\node[above] at (5.1,4) {$(r_1,2\pi)$};
				\draw[fill](4,4) circle(0.1cm);
				\node[below] at (12.5,-4) {$(1.3,0)$};
				\draw[fill](13,-4) circle(0.1cm);
				\node[above] at (13,4) {$(1.3,2\pi)$};
				\draw[fill](13,4) circle(0.1cm);
			\end{tikzpicture}
			
			\caption{\textbf{Visualization of the Shafranov and Czarny cross section geometries of a tokamak.}
				As both geometries can be described in curvilinear coordinates, mappings from a rectangular grid $[r_1,1.3]\times[0,2\pi]$ (center) to the considered geometries are indicated.
				The \textit{Shafranov} geometry (left) is given by the mapping $F_S:[r_1,1.3]\times[0,2\pi]\rightarrow \mathbb{R}^2$ and the \textit{Czarny} geometry (right) is given by the mapping $F_C:[r_1,1.3]\times[0,2\pi]\rightarrow \mathbb{R}^2$; see~\eqref{eq:shafranov} and~\eqref{eq:czarny}. 
				For the existence of the invertible mappings $F_S^{-1}$ and $F_C^{-1}$, we need $r_1>0$. 
				The depicted grids are symbolically refined at 2/3 of the generalized radius.
			}
			\label{fig:geom}
		\end{figure}
		
		In this paper, we consider three different geometries that represent cross sections of a tokamak; cf.~\cite{bouzat2018targeting,zoni2019solving,zoni2019theoretical}. 
		Two of the three geometries can be described by relatively short representations of curvilinear coordinates, i.e., based on a mapping $F_g$, $g\in\{S,C\}$, from the curvilinear coordinates $(r, \vartheta) \in [r_1, 1.3] \times [0, 2\pi)$, where $r$ is the (generalized) radius and $\vartheta$ the (generalized) angle to Cartesian coordinates $(x, y)$, defined as follows. 
		
		The {Shafranov} geometry is a deformed ellipse which is defined by the mapping
		\begin{align}
			\label{eq:shafranov}
			F_S(r,\vartheta):=\left(
			\begin{aligned}
				x(r, \vartheta) \\
				y(r, \vartheta)
			\end{aligned}
			\right)
			=
			\left(
			\begin{aligned}
				&x_0 + (1 - \kappa)r\,\cos\vartheta - \delta r^2 \\
				&y_0 + (1 + \kappa)r\,\sin\vartheta
			\end{aligned}
			\right);
		\end{align}
		see~\cref{fig:geom} (left). Here, $\kappa$ is the elongation and $\delta$ is the Shafranov shift; see~\cite{bouzat2018targeting,zoni2019solving}. For the parameters, we use $x_0=y_0=0$, $\kappa=0.3$, and $\delta=0.2$.
		
		The {Czarny} geometry is a D-shaped geometry and adds triangularity to the shape; see~\cref{fig:geom} (right). 
		It is defined by the mapping
		\begin{align}
			\label{eq:czarny}
			F_C(r,\vartheta):=\left(
			\begin{aligned}
				x(r, \vartheta) \\
				y(r, \vartheta) 
			\end{aligned}
			\right)
			=
			\left(
			\begin{aligned}
				&\frac{1}{\varepsilon} \left( 1 - \sqrt{1 + \varepsilon \left( \varepsilon + 2\,r\,\cos\vartheta \right)} \right) \\
				&y_0 + \frac{\textnormal{e}\, \xi\, r\,\sin\vartheta}{2 - \sqrt{1 + \varepsilon \left( \varepsilon + 2\,r\,\cos\vartheta \right)}}
			\end{aligned}    
			\right),
		\end{align}
		where $\varepsilon$ is the inverse aspect ratio, $\textnormal{e}$ the ellipticity, and $\xi=1/\sqrt{1 - \varepsilon^2/4}$,~see~\cite{CH08,zoni2019solving}. 
		For the parameters, we use $y_0 = 0$, $\varepsilon= 0.3$, and $\textnormal{e}= 1.4$.
		
		Note that for $r_1=0$, the inverse mappings $F_g^{-1}$, $g\in\{S,C\}$, do not exist as the functions $F_g$, $g\in\{S,C\}$, map the whole line $(0,\vartheta)$ on the origin. 
		This, however, is only of theoretical concerns as our method will use an interior radius $r_1$ such that $0<r_1\ll 1$.
		
		The third geometry, the nonanalytical \textit{Culham} geometry~\cite{connor1988tearing} has been chosen to take into account more realistic geometries in GYSELA. 
		The rather lengthy development can be found in~\cite[Sec. 6.2]{BLK23}. 
		For a visualization of the Culham cross sections in a 3D tokamak; see~\cref{fig:tokamak_cross_270}.

		\subsection{Discretization}\label{sec:disc}
		
		While a matrix-free implementation of iterative solvers based on finite element (FE) discretizations might be cumbersome, finite differences (FD) offer a straightforward approach for matrix-free implementations. 
		However, standard finite difference schemes applied to nonuniform meshes generally lead to nonsymmetric discretizations, even if the considered model problem is symmetric or, more precisely, the considered operator self-adjoint. 
		In~\cite{KKR21}, we presented a novel approach to derive symmetric FD discretizations for nonuniform meshes. 
		In this FE-inspired FD discretization, the energy functional is localized and discretized on the local elements.
		
		Let us first introduce a nonuniform mesh in product format by $r_1,\ldots,r_{n_r}$ with $0<r_1\ll 1$ and $r_{n_r}=1.3$ as well as $\vartheta_1,\ldots,\vartheta_{n_{\vartheta}+1}\in[0,2\pi]$ with $\vartheta_1=0$ and $\vartheta_{n_{\vartheta}+1}=2\pi$. We define
		\begin{align}
			\label{eq:definition_nr_ntheta}
			\begin{aligned}
				h_i := r_{i+1} - r_i ,\quad i\in\{1,\ldots,n_r-1\},\quad k_j := \vartheta_{j+1} - \vartheta_j,\quad j\in\{1,\ldots,n_{\vartheta}\}.
			\end{aligned}
		\end{align}
		
		Furthermore, we add an additional restriction for the discretization, i.e.,
		\begin{align}
			\label{eq:assumption_nr_ntheta}
			\begin{aligned}
				n_r\textnormal{ odd}, \quad &h_{2i} = h_{2i-1},\quad &i\in\{1,\ldots,(n_r-1)/2\},\\
				n_{\vartheta}\textnormal{ even}, \quad &k_{2j} = k_{2j-1},\quad &j\in\{1,\ldots,n_{\vartheta}/2\}.
			\end{aligned}
		\end{align}
		
		Note that this restriction is neither needed for the FD discretization nor for the geometric multigrid in general. 
		It will only be used to raise the convergence order of the geometric multigrid via implicit extrapolation. 
		For more details, see~\cite{KKR21,KKR22}.
		
		For theoretical purposes, we assume $h_i$ and $k_j$ to be uniformly bounded by
		\begin{align}
			0<h_{\min}\leq h_i \leq h\quad\text{and}\quad 0<k_{\min}\leq k_j\leq k,
		\end{align}
		as well as the existence of $0<\tau<\infty$ such that $h=\tau k$.
		
		We now localize $J(u)$ of~\eqref{eq:functional}, by considering the rectangular elements $R_{ij}:=[r_i,r_i+h_i]\times[\vartheta_j,\vartheta_j+k_j]$. 
		By transformation, the local energy functional writes
		\begin{align}
			\begin{aligned}
				J_{R_{i,j}}(u):&=\int_{R_{i,j}}\left(\frac{1}{2}\alpha|DF_g^{-T}\nabla_{(r,\vartheta)}u|^2+\frac{1}{2}\beta u^2-{f}{u}\right)|\det DF_g|\d(r,\vartheta),
			\end{aligned}
		\end{align}
		where $DF_g$ is the Jacobian matrix of $F_g$ and $DF_g^{-T}:=(DF_g^{T})^{-1}$, $g\in\{S,C\}$.
		
		For the sake of simplicity, we do not distinguish notations between functions defined on the logical and physical domain. 
		For a full derivation of the discretization, we refer to~\cite{KKR21,KKR22} with $\beta=0$ and~\cite{LSK25} for $\beta\neq 0$.
		In order to simplify the notation, we will furthermore use
		\begin{align}\label{eq:squarejacobian}
			\begin{aligned}
				\frac{1}{2}\alpha DF_g^{-1}DF_g^{-T}|\det DF_g|
				=:\begin{pmatrix}
					a^{rr} & \frac{1}{2}a^{r\vartheta}\\
					\frac{1}{2}a^{r\vartheta} & a^{\vartheta\vartheta}
				\end{pmatrix}.
			\end{aligned}
		\end{align}
		
		Discretizing the local energy components $J_{R_{i,j}}(u)$ to $\widetilde{J}_{R_{i,j}}(u)$, computing the derivative with respect to $u_{s,t}$, and searching for the critical point, i.e.,
		\begin{align}\label{eq:discretederivative}
			\begin{aligned}
				\sum_{i=1}^{n_{r}-1}\sum_{j=1}^{n_{\vartheta}}\frac{\partial}{\partial u_{s,t}}\widetilde{J}_{R_{i,j}}(u)\overset{!}{=}0.
			\end{aligned}
		\end{align}
		yields the nine-point finite difference stencil and right hand side as provided by~\cite{LSK25}. 
		Note that in the implementation, the geometric understanding has changed. 
		While this does not change the stencil itself, the particular interpretation of, e.g., left and top differ between the current codebase and~\cite{LSK25}. 
		
		In~\cite{KKR22}, we suggested a particular discretization \textit{across-the-origin} to handle the artificial singularity at $r_1=0$, avoiding to have the origin as a grid point. 
		It could be shown that this approach performed almost identical to incorporating (artificial) Dirichlet boundary conditions on an inner circle with generalized radius $0<r_1\ll 1$.
		
		\subsection{Multigrid and GMGPolar basics}
		
		GMGPolar is a geometric multigrid method that has been optimized to satisfy three desired requirements for the integration in a gyrokinetic framework such as GYSELA~\cite{grandgirard20165d}: 
		i) it achieves fast convergence for geometries represented in curvilinear or (generalized) polar coordinates, 
		ii) it provides a matrix-free approach with low memory requirements, and 
		iii) it realizes higher order convergence through implicit extrapolation. 
		As a multigrid method, it additionally allows for good parallel scalability by design. 
		In~\cite{LSK25}, it was shown that the number of floating point operations and memory cost of GMGPolar depend, asymptotically, only linearly on the number of degrees of freedom.
		In the following, we summarize very briefly GMGPolar's mathematical core properties.
		For more details and a pseudo-code description of GMGPolar, see~\cite{KKR22,LSK25}.
		
		\subsubsection{Smoothing for curvilinear coordinate representations}
		
		Through the transformation of the model problem~\eqref{eq:poisson} to a curvilinear coordinate representation, the ``strong connections'' between grid points change across the grid as we go from $r_1\approx 0$ to $r_{n_r}=1.3$. 
		This property strongly affects the choice for suitable smoothing operations in the multigrid algorithm. In general, point-wise smoothers are not sufficient.
		
		In~\cite{barros1988poisson}, smoothing properties of particular circular and radial line smoothers were considered analytically for polar coordinate representations. 
		Strong connections lie on circular lines near the origin, circle smoothers were found to perform better in this part of the disk-shaped domain. Closer to the boundary, strong connections lie on radial grid lines, and radial smoothers are more efficient.
		Based on these findings, GMGPolar switches from circle to radial smoothing where $k/h_ir_i>1$ is satisfied; a schematic picture is given in~\cref{fig:smoother_coloring_indices} (left). 
		Note that in most applications, we assume uniform discretization in the angular direction, i.e., $k=k_j$, $j=1,\ldots,n_{\vartheta}$. 
		Otherwise, a more general switching condition or overlapping smoothers are needed. 
		The implemented switching yields good smoothing behavior on the whole domain by only treating every grid point once per smoothing iteration.
		
		\subsubsection{Coarsening and intergrid transfer operators}
		
		Coarsening in GMGPolar is done by standard coarsening, selecting every second node in each dimension.
		Except for the particular handling of implicit extrapolation, as briefly mentioned in the next section, GMGPolar uses standard bilinear interpolation. 
		With the symmetry-preserving FD scheme, we can use the adjoint operator as restriction operator. 
		Note that we do not need any additional scaling constant between prolongation and restriction as our tailored FD scheme scales the right hand side locally with the surface of the considered rectangle; see also~\cite[Sec. 4.2]{KKR22}.
		
		\subsubsection{Higher order convergence through implicit extrapolation}
		
		When using the discretization from~\cref{sec:disc}, implicit extrapolation allows to raise the convergence order towards the true solution when refining the grid. 
		For the considered model problems and geometries, increases from order 2 to approximately 3.6 - 4.0 were observed~\cite{KKR21}. 
		For implicit extrapolation to take effect, GMGPolar makes some adjustments to standard multigrid algorithms.
		However, these changes only affect the finest two grid levels. 
		All operations between grid levels below the second finest grid comply with standard multigrid practices. 
		For a complete description of the adjusted intergrid transfer operators and smoothing operations, see~\cite[Sec. 4.3]{KKR22}.
		
		\subsubsection{Take and give approaches}
		
		A straightforward way to implement the FD stencil is the node-wise computation of all available row entries of the stiffness matrix. 
		This approach has been denoted the \textit{A-take}, or simply \textit{Take}, approach in~\cite{LSK25}, as it takes the necessary values from the memory locations of the neighboring nodes. 
		From the structure of the stencil, it can be seen that many entries are recomputed with the take approach. 
		Alternatively, we can optimize the computation of the stencil values per node and distribute computed values to the memory of neighboring nodes that need the same function evaluations. 
		This approach was denoted \textit{A-give} or \textit{Give}. 
		Both approaches will be discussed in the sections on the refactored version of GMGPolar.

		\section{Object-oriented redesign and algorithmic optimization}\label{sec:oore}
		The refactoring of GMGPolar represents a substantial transition from a functional programming style, which had been inherited from the initial implementation of~\cite{KKR22}, to a structured and object-oriented design. 
		With this shift, specific functionalities are now better encapsulated in dedicated classes, clarifying the responsibilities of different components within the codebase -- also ensuring better maintenance and extension capabilities. 
		The refactoring was essentially done during the master thesis of Julian Litz~\cite{Litz25}. Here, we outline the key aspects of the novel implementation.
		
		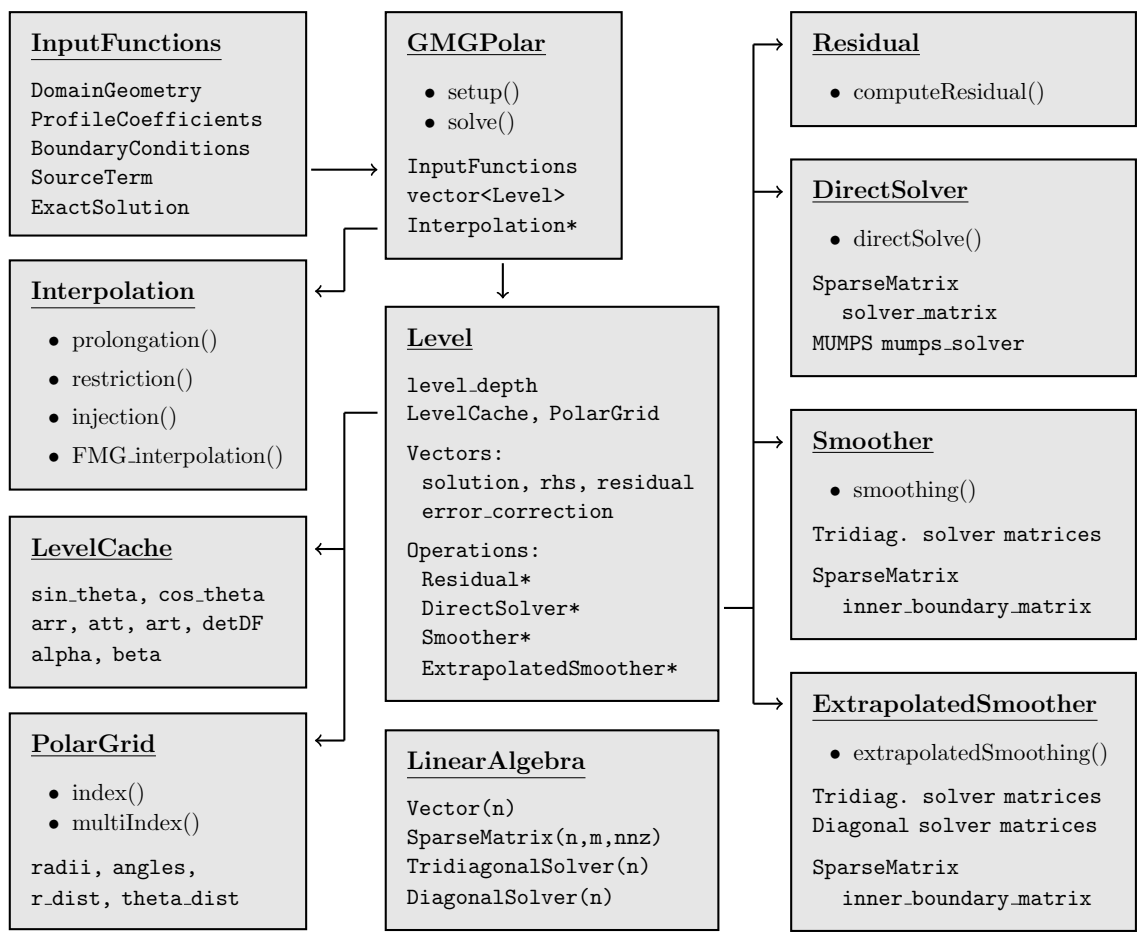
\begin{figure}[b!]
			\centering
			\scalebox{0.98}{
				\begin{tikzpicture}
					\node[
					draw=black, thick, fill=gray!20, 
					anchor=north west, text width=4.2cm, 
					align=left, inner sep=10pt,scale = 0.82] 
					at (0,0) 
					{
						\underline{\textbf{InputFunctions}}
						\small
						\\ \vspace{8pt}
						\texttt{DomainGeometry \\
							ProfileCoefficients \\
							BoundaryConditions \\
							SourceTerm \\
							ExactSolution
						}
					};
					
					\node[
					draw=black, thick, fill=gray!20, 
					anchor=north west, text width=3.2cm, 
					align=left, inner sep=10pt,scale = 0.82] 
					at (5.1,0) 
					{
						\underline{\textbf{GMGPolar}}
						\small
						\begin{itemize}
							\setlength{\itemindent}{-10pt}
							\setlength{\itemsep}{0pt} 
							\item setup()
							\vspace{-4pt}
							\item solve()
						\end{itemize}
						\vspace{-2pt}
						\texttt{InputFunctions \\
							vector<Level> \\
							Interpolation*
						}
					};
					
					\node[
					draw=black, thick, fill=gray!20, 
					anchor=north west, text width=5cm, 
					align=left, inner sep=10pt,scale = 0.82] 
					at (10.6,0) 
					{
						\underline{\textbf{Residual}}
						\small
						\begin{itemize}
							\setlength{\itemindent}{-10pt}
							\setlength{\itemsep}{0pt} 
							\item computeResidual()
						\end{itemize}
					};
					
					\node[
					draw=black, thick, fill=gray!20, 
					anchor=north west, text width=5cm, 
					align=left, inner sep=10pt,scale = 0.82] 
					at (10.6, -2.0) 
					{
						\underline{\textbf{DirectSolver}}
						\small
						\begin{itemize}
							\setlength{\itemindent}{-10pt}
							\setlength{\itemsep}{0pt} 
							\item directSolve()
						\end{itemize}
						\vspace{-2pt}
						\texttt{SparseMatrix 
							\hspace*{10 pt} solver\_matrix \\
							MUMPS mumps\_solver
						}
					};
					
					\node[
					draw=black, thick, fill=gray!20, 
					anchor=north west, text width=5.0cm, 
					align=left, inner sep=10pt,scale = 0.82] 
					at (10.6, -5.4) 
					{
						\underline{\textbf{Smoother}}
						\small
						\begin{itemize}
							\setlength{\itemindent}{-10pt}
							\setlength{\itemsep}{0pt} 
							\item smoothing()
						\end{itemize}
						\vspace{-2pt}
						\texttt{Tridiag. solver matrices \\
							\vspace{5pt}
							SparseMatrix 
							\hspace*{10 pt} inner\_boundary\_matrix
						}
					};
					
					\node[
					draw=black, thick, fill=gray!20, 
					anchor=north west, text width=5.0cm, 
					align=left, inner sep=10pt,scale = 0.82] 
					at (10.6, -8.97) 
					{
						\underline{\textbf{ExtrapolatedSmoother}}
						\small
						\begin{itemize}
							\setlength{\itemindent}{-10pt}
							\setlength{\itemsep}{0pt} 
							\item extrapolatedSmoothing()
						\end{itemize}
						\vspace{-2pt}
						\texttt{Tridiag. solver matrices \\
							Diagonal solver matrices \\
							\vspace{5pt}
							SparseMatrix 
							\hspace*{10 pt} inner\_boundary\_matrix
						}
					};
					
					\node[
					draw=black, thick, fill=gray!20, 
					anchor=north west, text width=4.8cm, 
					align=left, inner sep=10pt,scale = 0.82] 
					at (5.1, -4.0) 
					{
						\underline{\textbf{Level}}
						\small
						\\ \vspace{8pt}
						\texttt{level\_depth \\
							LevelCache, PolarGrid \\ 
							\vspace{5pt}
							Vectors:\\
							\hspace{7pt}solution, rhs, residual \\
							\hspace{7pt}error\_correction\\
							\vspace{5pt}
							Operations: \\
							\hspace{7pt}Residual* \\
							\hspace{7pt}DirectSolver*\\
							\hspace{7pt}Smoother* \\
							\hspace{7pt}ExtrapolatedSmoother* 
						}
					};
					
					\node[
					draw=black, thick, fill=gray!20, 
					anchor=north west, text width=4.2cm, 
					align=left, inner sep=10pt,scale = 0.82] 
					at (0,-3.37) 
					{
						\underline{\textbf{Interpolation}}
						\small
						\begin{itemize}
							\setlength{\itemindent}{-10pt}
							\setlength{\itemsep}{0pt} 
							\item prolongation()
							\item restriction()
							\item injection()
							\item FMG\_interpolation()
						\end{itemize}
					};
					
					\node[
					draw=black, thick, fill=gray!20, 
					anchor=north west, text width=4.8cm, 
					align=left, inner sep=10pt,scale = 0.82] 
					at (5.1,-9.75) 
					{
						\underline{\textbf{LinearAlgebra}}
						\small
						\\ \vspace{8pt}
						\texttt{Vector(n) \\
							SparseMatrix(n,m,nnz) \\
							TridiagonalSolver(n) \\
							DiagonalSolver(n)
						}
					};
					
					\node[
					draw=black, thick, fill=gray!20, 
					anchor=north west, text width=4.2cm, 
					align=left, inner sep=10pt,scale = 0.82] 
					at (0, -6.85) 
					{
						\underline{\textbf{LevelCache}}
						\small
						\\ \vspace{8pt}
						\texttt{sin\_theta, cos\_theta \\
							arr, att, art, detDF \\
							alpha, beta
						}
					};
					
					\node[
					draw=black, thick, fill=gray!20, 
					anchor=north west, text width=4.2cm, 
					align=left, inner sep=10pt,scale = 0.82] 
					at (0,-9.52)
					{
						\underline{\textbf{PolarGrid}}
						\small
						\begin{itemize}
							\setlength{\itemindent}{-10pt}
							\setlength{\itemsep}{0pt} 
							\item index()
							\vspace{-4pt}
							\item multiIndex()
						\end{itemize}
						\vspace{-2pt}
						\texttt{radii, angles, \\
							r\_dist, theta\_dist
						}
					};

					\draw[->, thick] (4.1, -2.15) -- (5.0, -2.15);
					
					\draw[thick] (5.0, -2.95) -- (4.55, -2.95);
					\draw[thick] (4.55, -2.95) -- (4.55, -3.8);
					\draw[->, thick] (4.55, -3.8) -- (4.15, -3.8);
					
					\draw[->, thick] (6.7, -3.42) -- (6.7, -3.9);
					
					\draw[thick] (5.0, -5.45) -- (4.55, -5.45);
					\draw[thick] (4.55, -5.45) -- (4.55, -9.9);
					\draw[->, thick] (4.55, -9.9) -- (4.15, -9.9);
					\draw[->, thick] (4.55, -7.3) -- (4.15, -7.3);
					
					\draw[thick] (9.7, -8.1) -- (10.1, -8.1);
					\draw[thick] (10.1, -9.4) -- (10.1, -0.45);
					\draw[->, thick] (10.1, -0.45) -- (10.5, -0.45);
					\draw[->, thick] (10.1, -2.45) -- (10.5, -2.45);
					\draw[->, thick] (10.1, -5.85) -- (10.5, -5.85);
					\draw[->, thick] (10.1, -9.4) -- (10.5, -9.4);

				\end{tikzpicture}
			}
			\caption{\textbf{Refactored class layout highlighting the modular structure and object-oriented design approach.}}
			\label{fig:refactored_layout}
		\end{figure}
		
		Compared to the previous version 1 of GMGPolar, the data structure has been separated into more specialized classes, instead of being inside a unique large multigrid \textit{Level} class. 
		A dedicated \texttt{PolarGrid} class will manage grid-related data, ensuring efficient access and organization, while a separate \texttt{LevelCache} class will handle the storage and retrieval of precomputed data. 
		The \texttt{Interpolation} class manages intergrid transfer operations, separating the responsibilities of data movement between grid levels from other functionalities of the solver. 
		In addition, a custom \texttt{LinearAlgebra} class manages the fundamental operations on vectors and matrices. 
		In this class, we implemented tailored tridiagonal solver algorithms, which are crucial for the performance of the smoother.
		
		To further modularize the design, distinct operator classes were created to handle specific computational tasks, such as computing the \texttt{Residual}, solving the system coarse matrix via \texttt{DirectSolver}, and performing smoothing operations via \texttt{Smoother}.
		The refactored layout is presented in~\cref{fig:refactored_layout}. 
		For an overview of particular settings with respect to geometry, multigrid, and particular problem settings, see~\cref{tab:parameters}. 
		By adopting this more object-oriented approach, which also optimizes memory usage and enhances the multigrid cycle methodology, we obtain a more flexible and efficient tool for large-scale gyrokinetic simulations.

		The original GMGPolar code employed exclusively the Give approach for matrix-free computations and dynamically computed transformation coefficients during matrix-vector operations, whereas the Take approach was only available with the assembled matrix version.
		While the Give method minimizes memory usage, it introduces a computational overhead, particularly for complex geometries. 
		To provide greater flexibility, the refactored implementation supports, both, the Give and Take stencil implementations, allowing to choose between storing or recomputing transformation coefficients $a^{rr}$, $a^{r \vartheta}$, $a^{\vartheta \vartheta}$ and $\det DF_g$ from~\eqref{eq:squarejacobian} based on specific requirements. 
		With this functionality, the solver adapts to a wider range of problem sizes and computational constraints. 
		In the following, we will present in depth the improvements that were undertaken in GMGPolar version 2.
		
		\subsection{Memory usage and cache efficiency}\label{sec:cache}
		
		Memory and cache optimization was a key component to be considered in the refactoring phase. 
		The matrix-free GMGPolar implementation had already been designed with small memory requirements~\cite{LSK25} compared to other state-of-the-art solvers~\cite{BLK23}. 
		Nevertheless, the previous implementation did not yet exploit the symmetry of the smoother matrices, relying instead on a full LU decomposition that introduced additional fill-in. 
		While the total memory consumption for the prior version was computed asymptotically linear as $12n$~\cite{LSK25} (accounting for all multigrid levels), the new Give variant reduces finest-level storage to just $5n$ (asymptotically with all levels: $6.7n$) by employing in-place symmetric Cholesky factorizations for the smoothers and eliminating redundant temporary vectors.
		Here, $n = n_r \cdot n_\vartheta$ denotes the total number of grid nodes. The Take variant introduces four extra vectors (for arrays \verb|arr|, \verb|att|, \verb|art|, and \verb|detDF|) to streamline data access, bringing its peak requirement to $9n$ (asymptotically with all levels: $12n$).

		
		The smoother constitutes a major computational component of the solver (see~\cite{LSK25}); therefore, both the setup and solution phases of the smoother matrices were improved. In the previous version, the assembled matrices were stored in Coordinate List (COO) format and decomposed using a general LU decomposition.
		The smoother's solver matrices naturally take the form of tridiagonal matrices or cyclic tridiagonal matrices when using circle line smoothing with peridodic boundary conditions; cf.~\cite{KKR22}. 
		In the new implementation, we eliminate explicit row–column indices and exploit the symmetry, only storing the upper or lower half of the matrices and additionally use specialized tridiagonal solvers suited to this structure.
		The tridiagonal matrices of the radial smoother are factorized using Cholesky decomposition in $4n + \mathcal{O}(1)$ operations and solved via forward and backward substitution in $5n + \mathcal{O}(1)$ operations.
		To factorize the cyclic tridiagonal matrices of the circle smoother, we apply the Sherman–Morrison formula, which  expresses the cyclic tridiagonal matrix as a rank-one modification of a standard tridiagonal matrix, enabling efficient factorization without fill-in. The solution phase involves solving two independent tridiagonal systems and combining their results, requiring $12n + \mathcal{O}(1)$ operations.
		Although this new approach incurs a slightly higher cost in the solution phase compared to the adapted LU decomposition used previously (cf.~\cite[Sec. 8.4]{LSK25}), it avoids the fill-in introduced in the last row and column, reducing memory footprint. 
		
\begin{figure}
			\centering
			\raisebox{+0.1\height}{
				\begin{tikzpicture}[scale=2.2, transform shape]
					\pgfmathsetmacro{\a}{360 / 32}
					\pgfmathsetmacro{\r}{1}
					
					\draw (0,0) circle (\r cm);
					\fill[black!60] (0,0) -- (0*\a: \r cm) arc[start angle = 0*\a, end angle=1*\a, radius=\r cm] -- cycle;
					\fill[black!60] (0,0) -- (2*\a: \r cm) arc[start angle = 2*\a, end angle=3*\a, radius=\r cm] -- cycle;
					\fill[black!60] (0,0) -- (4*\a: \r cm) arc[start angle = 4*\a, end angle=5*\a, radius=\r cm] -- cycle;
					\fill[black!60] (0,0) -- (6*\a: \r cm) arc[start angle = 6*\a, end angle=7*\a, radius=\r cm] -- cycle;
					\fill[black!60] (0,0) -- (8*\a: \r cm) arc[start angle = 8*\a, end angle=9*\a, radius=\r cm] -- cycle;
					\fill[black!60] (0,0) -- (10*\a: \r cm) arc[start angle = 10*\a, end angle=11*\a, radius=\r cm] -- cycle;
					\fill[black!60] (0,0) -- (12*\a: \r cm) arc[start angle = 12*\a, end angle=13*\a, radius=\r cm] -- cycle;
					\fill[black!60] (0,0) -- (14*\a: \r cm) arc[start angle = 14*\a, end angle=15*\a, radius=\r cm] -- cycle;
					\fill[black!60] (0,0) -- (16*\a: \r cm) arc[start angle = 16*\a, end angle=17*\a, radius=\r cm] -- cycle;
					\fill[black!60] (0,0) -- (18*\a: \r cm) arc[start angle = 18*\a, end angle=19*\a, radius=\r cm] -- cycle;
					\fill[black!60] (0,0) -- (20*\a: \r cm) arc[start angle = 20*\a, end angle=21*\a, radius=\r cm] -- cycle;
					\fill[black!60] (0,0) -- (22*\a: \r cm) arc[start angle = 22*\a, end angle=23*\a, radius=\r cm] -- cycle;
					\fill[black!60] (0,0) -- (24*\a: \r cm) arc[start angle = 24*\a, end angle=25*\a, radius=\r cm] -- cycle;
					\fill[black!60] (0,0) -- (26*\a: \r cm) arc[start angle = 26*\a, end angle=27*\a, radius=\r cm] -- cycle;
					\fill[black!60] (0,0) -- (28*\a: \r cm) arc[start angle = 28*\a, end angle=29*\a, radius=\r cm] -- cycle;
					\fill[black!60] (0,0) -- (30*\a: \r cm) arc[start angle = 30*\a, end angle=31*\a, radius=\r cm] -- cycle;
					
					\pgfmathsetmacro{\l}{8 / 16} 
					\pgfmathsetmacro{\N}{8}
					\foreach \i in {1,...,\N} {
						\pgfmathsetmacro{\radius}{\l * ((\N - \i + 1) / \N)}
						\ifodd\i
						\fill[black!60] (0,0) circle (\radius);
						\else
						\fill[white] (0,0) circle (\radius);
						\fi
					}

					\pgfmathsetmacro{\l}{1}
					\pgfmathsetmacro{\N}{16}
					\pgfmathsetmacro{\nl}{15}
					
					\pgfmathsetmacro{\a}{360 / 32}
					\foreach \anglei in {0,...,31} {
						\pgfmathsetmacro{\angle}{\anglei * 360 / 32}
						\draw[thin] (\angle: 3 * 1.0 / 16) -- (\angle:1.0);
						
						\draw[very thin, black!100] (\angle: 1 * 1.0 / 16) -- (\angle: 3 * 1.0 / 16);
						\draw[ultra thin, black!100] (\angle: 0.012cm) -- (\angle: 1 * 1.0 / 16);
					}

					\draw[white, line width=0.15mm] (0,0) circle (0.509 cm);
					
					\foreach \i in {1,...,\nl} {
						\pgfmathsetmacro{\radius}{\l * ((\N - \i + 1) / \N)}
						\draw[thin] (0,0) circle (\radius);
					}

					\draw[thin] (0,0) circle (\l / \N);
					\draw[ultra thin] (0,0) circle (0.012cm);
					
					\draw[red!100!black, dashed, line width=0.5mm, dash pattern=on 6pt off 3pt] (0,0) circle (0.509 cm);

				\end{tikzpicture}
			}
			\hspace*{0.5cm}
			\raisebox{-0.0\height}{
				\begin{tikzpicture}[scale=0.95, transform shape]
					\scriptsize
					
					\draw[gray, line width=0.5mm] (0,0) -- (7,0);
					\draw[gray, line width=0.5mm] (0,1) -- (7,1);
					\draw[gray, line width=0.5mm] (0,2) -- (7,2);
					
					\draw[gray, line width=0.5mm] (0,0) -- (0,3);
					\draw[gray, line width=0.5mm] (1,0) -- (1,3);
					\draw[gray, line width=0.5mm] (2,0) -- (2,3);
					\draw[gray, line width=0.5mm] (3,0) -- (3,3);
					\draw[gray, line width=0.5mm] (4,0) -- (4,3);
					\draw[gray, line width=0.5mm] (5,0) -- (5,3);
					\draw[gray, line width=0.5mm] (6,0) -- (6,3);
					\draw[gray, line width=0.5mm] (7,0) -- (7,3);

					\draw[violet!75!black, ->, line width=0.5mm] (0,0) -- (0,-0.65);
					\draw[violet!75!black, ->, line width=0.5mm] (1,0) -- (1,-0.65);
					\draw[violet!75!black, ->, line width=0.5mm] (2,0) -- (2,-0.65);
					\draw[violet!75!black, ->, line width=0.5mm] (3,0) -- (3,-0.65);
					\draw[violet!75!black, ->, line width=0.5mm] (4,0) -- (4,-0.65);
					\draw[violet!75!black, ->, line width=0.5mm] (5,0) -- (5,-0.65);
					\draw[violet!75!black, ->, line width=0.5mm] (6,0) -- (6,-0.65);
					\draw[violet!75!black, ->, line width=0.5mm] (7,0) -- (7,-0.65);
					
					\draw[violet!75!black, ->, line width=0.5mm] (0,3) -- (0,3.65);
					\draw[violet!75!black, ->, line width=0.5mm] (1,3) -- (1,3.65);
					\draw[violet!75!black, ->, line width=0.5mm] (2,3) -- (2,3.65);
					\draw[violet!75!black, ->, line width=0.5mm] (3,3) -- (3,3.65);
					\draw[violet!75!black, ->, line width=0.5mm] (4,3) -- (4,3.65);
					\draw[violet!75!black, ->, line width=0.5mm] (5,3) -- (5,3.65);
					\draw[violet!75!black, ->, line width=0.5mm] (6,3) -- (6,3.65);
					\draw[violet!75!black, ->, line width=0.5mm] (7,3) -- (7,3.65);

					\node[circle, draw=black, fill=white, text=black, minimum size=0.5cm, inner sep=2pt] at (0,0) {0};
					\node[circle, draw=black, fill=white, text=black, minimum size=0.5cm, inner sep=2pt] at (0,1) {1};
					\node[circle, draw=black, fill=white, text=black, minimum size=0.5cm, inner sep=2pt] at (0,2) {2};
					\node[circle, draw=black, fill=white, text=black, minimum size=0.5cm, inner sep=2pt] at (0,3) {3};
					
					\node[circle, draw=black, fill=black, text=white, minimum size=0.5cm, inner sep=2pt] at (1,0) {4};
					\node[circle, draw=black, fill=black, text=white, minimum size=0.5cm, inner sep=2pt] at (1,1) {5};
					\node[circle, draw=black, fill=black, text=white, minimum size=0.5cm, inner sep=2pt] at (1,2) {6};
					\node[circle, draw=black, fill=black, text=white, minimum size=0.5cm, inner sep=2pt] at (1,3) {7};
					
					\node[circle, draw=black, fill=white, text=black, minimum size=0.5cm, inner sep=2pt] at (2,0) {8};
					\node[circle, draw=black, fill=white, text=black, minimum size=0.5cm, inner sep=2pt] at (2,1) {9};
					\node[circle, draw=black, fill=white, text=black, minimum size=0.5cm, inner sep=2pt] at (2,2) {10};
					\node[circle, draw=black, fill=white, text=black, minimum size=0.5cm, inner sep=2pt] at (2,3) {11};
					
					\node[circle, draw=black, fill=black, text=white, minimum size=0.5cm, inner sep=2pt] at (3,0) {12};
					\node[circle, draw=black, fill=black, text=white, minimum size=0.5cm, inner sep=2pt] at (3,1) {13};
					\node[circle, draw=black, fill=black, text=white, minimum size=0.5cm, inner sep=2pt] at (3,2) {14};
					\node[circle, draw=black, fill=black, text=white, minimum size=0.5cm, inner sep=2pt] at (3,3) {15};
					
					\node[circle, draw=black, fill=black, text=white, minimum size=0.5cm, inner sep=2pt] at (4,0) {16};
					\node[circle, draw=black, fill=black, text=white, minimum size=0.5cm, inner sep=2pt] at (5,0) {17};
					\node[circle, draw=black, fill=black, text=white, minimum size=0.5cm, inner sep=2pt] at (6,0) {18};
					\node[circle, draw=black, fill=black, text=white, minimum size=0.5cm, inner sep=2pt] at (7,0) {19};
					
					\node[circle, draw=black, fill=white, text=black, minimum size=0.5cm, inner sep=2pt] at (4,1) {20};
					\node[circle, draw=black, fill=white, text=black, minimum size=0.5cm, inner sep=2pt] at (5,1) {21};
					\node[circle, draw=black, fill=white, text=black, minimum size=0.5cm, inner sep=2pt] at (6,1) {22};
					\node[circle, draw=black, fill=white, text=black, minimum size=0.5cm, inner sep=2pt] at (7,1) {23};
					
					\node[circle, draw=black, fill=black, text=white, minimum size=0.5cm, inner sep=2pt] at (4,2) {24};
					\node[circle, draw=black, fill=black, text=white, minimum size=0.5cm, inner sep=2pt] at (5,2) {25};
					\node[circle, draw=black, fill=black, text=white, minimum size=0.5cm, inner sep=2pt] at (6,2) {26};
					\node[circle, draw=black, fill=black, text=white, minimum size=0.5cm, inner sep=2pt] at (7,2) {27};
					
					\node[circle, draw=black, fill=white, text=black, minimum size=0.5cm, inner sep=2pt] at (4,3) {28};
					\node[circle, draw=black, fill=white, text=black, minimum size=0.5cm, inner sep=2pt] at (5,3) {29};
					\node[circle, draw=black, fill=white, text=black, minimum size=0.5cm, inner sep=2pt] at (6.,3) {30};
					\node[circle, draw=black, fill=white, text=black, minimum size=0.5cm, inner sep=2pt] at (7,3) {31};
					
					\draw[red!100!black, dashed, line width=0.5mm, dash pattern=on 6pt off 3pt] (3.5,-0.8) -- (3.5,3.8);
					
					\node[red!100!black] at (4.5,-1.2) {\footnotesize Separation of the smoother sections};

					\draw[gray, ->, line width=0.7mm]
					(-1,-1.2) -- (-1,1) node[pos=1, above, text=black] {\footnotesize $\vartheta$};
					
					\draw[gray, ->, line width=0.7mm]
					(-1,-1.2) -- (1,-1.2) node[pos=1, right, text=black] {\footnotesize $r$};

				\end{tikzpicture}
			}
			\caption{\textbf{Combined circle-radial smoother and indexing for the smoothing operations.} 
				Circle and radial lines colored black and white on a grid of dimension \(16 \times 32\) with eight circle lines of 32 nodes and 16 radial lines of eight nodes; for visualization simplification, the curvilinear lines are shown without the nodes (left). 
				Optimized grid indexing for a periodic grid of dimension \(8 \times 4\) with four circle lines of four nodes and four radial lines of four nodes. 
				Vertical lines of the same color represent circular smoothers, while horizontal lines of the same color correspond to radial smoothers (right).
			}
			\label{fig:smoother_coloring_indices}
		\end{figure}
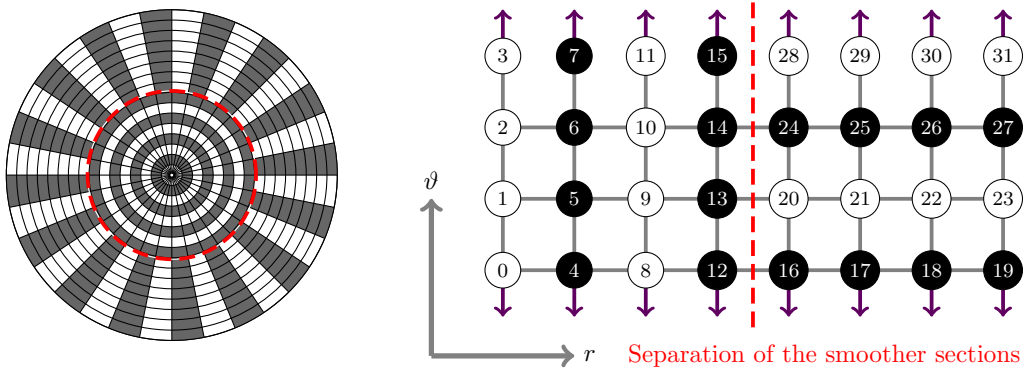

		Moreover, we optimized the solution step of the smoother matrices on cache level by reordering grid indexing to align with the smoother's line patterns. 
		For a visualization of a smoother-aligned indexing, see~\cref{fig:smoother_coloring_indices}. 
		This reorganization increases data locality and improves cache line usage, consequently, leading to faster execution times.
		
		\begin{remark}
			For solving the circle smoothing system on the innermost circle with the default, across-the-origin, discretization, attention has to be paid. Due to the across-the-origin approach, the corresponding submatrix is not tridiagonal. Therefore, we use the COO format and MUMPS for the smoother on the innermost circle.
		\end{remark}
		
		As sparse linear algebra applications are often memory-bound, improving memory access patterns is also promising for speeding up the particular application. 
		However, as first observed in~\cite{BLK23}, the matrix-free version of GMGPolar, replacing most memory accesses by (re)computations, did not speed up our method as expected. 
		This observation is a direct consequence of the complex domain geometries of the tokamak cross sections. 
		On these geometries, the evaluation of the sine and cosine functions as well as the factors $a^{rr}, a^{r\vartheta}, a^{\vartheta\vartheta}$ were found to be relatively costly. 
		In~\cite{LSK25}, we already stored sine and cosine evaluations for the different generalized angles. 
		In the novel version, these values are also stored. By default, the novel version also caches the values of $\alpha$ and $\beta$ evaluations from~\eqref{eq:poisson} and~\eqref{eq:alpha_zoni} for the Give approach. 
		For the Take approach, we additionally cache the transformation coefficients from~\eqref{eq:squarejacobian} -- which can also be stored for the Give approach upon selection by the user. 
		
		
		\subsection{Parallelization}\label{sec:parallel} 
		
		In the original implementation of~\cite{LSK25}, a task-based parallelism with dependencies was used. 
		At runtime, the OpenMP threads could pick up and perform tasks as they became available. 
		While this approach might be advantageous for unstructured task sets and largely differently sized task, we found that a loop-based parallelism yielded better results for the structured and similarly-sized problems considered in GMGPolar. 
		With the loop-based parallelism, we avoid dynamic scheduling and dependency management and reduce synchronization overhead.
		
\begin{figure}[h]
			\centering
			\begin{tikzpicture}[scale=0.9]
				\begin{scope}[xshift=0cm, yshift=0cm, scale=0.7] 
					\draw (0,0) circle[radius=10pt];
					\node at (0,0) {$0$};
					\draw (4.5,0) circle[radius=10pt];
					\node at (4.5,0) {$3$};
					\draw (9.0,0) circle[radius=10pt];
					\node at (9.0,0) {$6$};
					\draw (9.0,0) circle[radius=10pt];
					\node at (9.0,0) {$6$};
					
					\draw (1.5,-2) circle[radius=10pt];
					\node at (1.5,-2) {$1$};
					\draw (6.0,-2) circle[radius=10pt];
					\node at (6.0,-2) {$4$};
					\draw (10.5,-2) circle[radius=10pt];
					\node at (10.5,-2) {$7$};
					
					\draw (3.0,-4) circle[radius=10pt];
					\node at (3.0,-4) {$2$};
					\draw (7.5,-4) circle[radius=10pt];
					\node at (7.5,-4) {$5$};
					\draw (12.0,-4) circle[radius=10pt];
					\node at (12.0,-4) {$8$};
					
					\draw[->, thick] (1.5 - 0.3, -2 + 0.4) -- (0 + 0.3, 0 - 0.4);
					\draw[->, thick] (1.5 + 4.5 - 0.3, -2 + 0.4) -- (0 + 4.5 + 0.3, 0 - 0.4);
					\draw[->, thick] (1.5 + 9.0 - 0.3, -2 + 0.4) -- (0 + 4.5 + 4.5 + 0.3, 0 - 0.4);
					\draw[->, thick] (1.5 + 1.5 - 0.3, -4 + 0.4) -- (0 + 1.5 + 0.3, -2 - 0.4);
					\draw[->, thick] (1.5 + 4.5 + 1.5 - 0.3, -4 + 0.4) -- (0 + 1.5 + 4.5 + 0.3, -2 - 0.4);
					\draw[->, thick] (1.5 + 9.0 + 1.5 - 0.3, -4 + 0.4) -- (0 + 1.5 + 4.5 + 4.5 + 0.3, -2 - 0.4);

					\draw[->, thick] (1.5 + 0.4, -2 + 0.3) -- (4.5 - 0.4, 0 - 0.3);
					\draw[->, thick] (3.0 + 0.4, -4 + 0.3) -- (6.0 - 0.4, -2 - 0.3);
					
					\draw[->, thick] (1.5 + 4.5 + 0.4, -2 + 0.3) -- (4.5 + 4.5 - 0.4, 0 - 0.3);
					\draw[->, thick] (3.0 + 4.5 + 0.4, -4 + 0.3) -- (6.0 + 4.5 - 0.4, -2 - 0.3);
					
				\end{scope}
			\end{tikzpicture}
			\caption{\textbf{Dependency graph for the application of the system matrix $A^{(i)}$, of multigrid level $i\in\{0,\ldots,L-1\}$, using the Give implementation.} Each vertex represents a task corresponding to a line of nodes. The edges visualize the dependencies between these tasks.}
			\label{fig:applyA_parallelization}
		\end{figure}
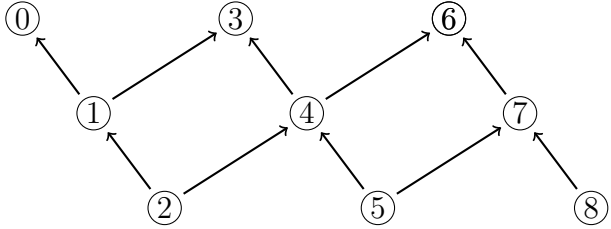

		In order to avoid concurrent updates of memory locations when applying the system matrix $A$ with the Give approach, GMGPolar treats every third line in parallel; see~\cite{LSK25} and~\cref{fig:applyA_parallelization}.
		
\begin{figure}[h!]
			\begin{adjustwidth}{-.75in}{0in}
				\centering
				\begin{tikzpicture}[scale=0.85]
					\begin{scope}[xshift=0cm, yshift=0cm, scale=0.7] 
						\draw[->, thick] (1.5 + 0.75 - 0.1, -3.5 + 0.5) -- (1.5 + 0.2,0 - 0.5);
						\draw[->, thick] (1.5 + 0.75 + 0.1, -3.5 + 0.5) -- (3.0 - 0.2,0 - 0.5);
						\draw[->, thick] (4.5 + 0.75 - 0.1, -3.5 + 0.5) -- (4.5 + 0.2,0 - 0.5);
						\draw[->, thick] (4.5 + 0.75 + 0.1, -3.5 + 0.5) -- (6.0 - 0.2,0 - 0.5);
						
						\draw[->, thick] (1.5 - 0.1, -11.0 + 0.5) -- (1.5 - 0.75 + 0.2,-7.5 - 0.5);
						\draw[->, thick] (1.5 + 0.1, -11.0 + 0.5) -- (3.0 - 0.75 - 0.2,-7.5 - 0.5);
						\draw[->, thick] (4.5 - 0.1, -11.0 + 0.5) -- (4.5 - 0.75 + 0.2,-7.5 - 0.5);
						\draw[->, thick] (4.5 + 0.1, -11.0 + 0.5) -- (6.0 - 0.75 - 0.2,-7.5 - 0.5);

						\draw[->, thick] (1.5 + 0.75 + 9.5 - 0.1, -3.5 + 0.5) -- (1.5 + 9.5 + 0.2,0 - 0.5);
						\draw[->, thick] (1.5 + 0.75 + 9.5 + 0.1, -3.5 + 0.5) -- (3.0 + 9.5 - 0.2,0 - 0.5);
						\draw[->, thick] (4.5 + 0.75 + 9.5 + 0.1, -3.5 + 0.5) -- (4.5 + 9.5 + 0.2,0 - 0.5);
						\draw[->, thick] (4.5 + 0.75 + 9.5 - 0.5, -3.5 + 0.4) -- (0.0 + 9.5 + 0.3,0 - 0.5);
						
						\draw[->, thick] (1.5 + 9.5 - 0.1, -11.0 + 0.5) -- (1.5 - 0.75 + 0.2+ 9.5,-7.5 - 0.5);
						\draw[->, thick] (1.5+ 9.5 + 0.1, -11.0 + 0.5) -- (3.0 - 0.75 - 0.2+ 9.5,-7.5 - 0.5);
						\draw[->, thick] (4.5+ 9.5 - 0.1, -11.0 + 0.5) -- (4.5 - 0.75 + 0.2+ 9.5,-7.5 - 0.5);
						\draw[->, thick] (4.5+ 9.5 + 0.1, -11.0 + 0.5) -- (6.0 - 0.75 - 0.2+ 9.5,-7.5 - 0.5);

						\draw [decorate, decoration={brace, mirror, amplitude=5pt}] (-6, 0.5) -- (-6,-6);
						\node[align=center, text width=3.0cm] at (-7cm, -3cm) {\rotatebox[origin=c]{90}{\small \textbf{Color: Black}}};
						
						\draw [decorate, decoration={brace, mirror, amplitude=5pt}] (-0.75,+0.5) -- (-0.75,-4.5);
						\node[align=center, text width=3cm] at (-3.5cm, -2cm) {\small \textbf{Apply \( \mathbf{A_{s_cs_c^\perp}} \)}};
						
						\draw [decorate, decoration={brace, mirror, amplitude=5pt}] (-0.75, -5) -- (-0.75,-6);
						\node[align=center, text width=3cm] at (-3.5cm, -5.5cm) {\small \textbf{Solve system}};
						
						\draw [decorate, decoration={brace, mirror, amplitude=5pt}] (-6, -7) -- (-6,-13.5);
						\node[align=center, text width=3.0cm] at (-7cm, -10cm) {\rotatebox[origin=c]{90}{\small \textbf{Color: White}}};
						
						\draw [decorate, decoration={brace, mirror, amplitude=5pt}] (-0.75, - 7) -- (-0.75,-12);
						\node[align=center, text width=3cm] at (-3.5cm, -9.5cm) {\small \textbf{Apply \( \mathbf{A_{s_cs_c^\perp}} \)}};
						
						\draw [decorate, decoration={brace, mirror, amplitude=5pt}] (-0.75, -12.5) -- (-0.75,-13.5);    
						\node[align=center, text width=3.0cm] at (-3.5cm, -13cm) {\small \textbf{Solve system}};
						
						\fill[black] (0,0) circle[radius=10pt];
						\node[text=white] at (0,0) {$8$};
						\fill[black] (1.5,0) circle[radius=10pt];
						\node[text=white] at (1.5,0) {$6$};
						\fill[black] (3,0) circle[radius=10pt];
						\node[text=white] at (3,0) {$4$};
						\fill[black] (4.5,0) circle[radius=10pt];
						\node[text=white] at (4.5,0) {$2$};
						\fill[black] (6,0) circle[radius=10pt];
						\node[text=white] at (6,0) {$0$};
						
						\draw (0 + 0.75, -2) circle[radius=10pt];
						\node at (0 + 0.75, -2) {$7$};
						\draw (3 + 0.75, -2) circle[radius=10pt];
						\node at (3 + 0.75, -2) {$3$};
						\draw (6 + 0.75, -2) circle[radius=10pt];
						\node at (6 + 0.75, -2) {$\text{-}1$};
						
						\draw[->, thick] (0.75 - 0.1, -1.5) -- (0 + 0.1, -0.5);
						\draw[->, thick] (0.75 + 0.1, -1.5) -- (1.5 - 0.1, -0.5);
						\draw[->, thick] (0.75+3 - 0.1, -1.5) -- (3 + 0.1, -0.5);
						\draw[->, thick] (0.75+3 + 0.1, -1.5) -- (4.5 - 0.1, -0.5);
						\draw[->, thick] (0.75+6 - 0.1, -1.5) -- (6 + 0.1, -0.5);
						
						\draw (1.5 + 0.75, -3.5) circle[radius=10pt];
						\node at (1.5 + 0.75, -3.5) {$5$};
						\draw (4.5 + 0.75, -3.5) circle[radius=10pt];
						\node at (4.5 + 0.75, -3.5) {$1$};
						
						\draw[->, dashed, thick] (1.5 + 0.75 -0.4, -3.5 + 0.3) -- (0 + 0.75 + 0.3, -2 -0.4);
						\draw[->, dashed, thick] (1.5 + 0.75 +0.4, -3.5 + 0.3) -- (3 + 0.75 - 0.3, -2 -0.4);
						\draw[->, dashed, thick] (1.5+3 + 0.75 -0.4, -3.5 + 0.3) -- (3 + 0.75 + 0.3, -2 -0.4);
						\draw[->, dashed, thick] (1.5+3 + 0.75 +0.4, -3.5 + 0.3) -- (6 + 0.75 - 0.3, -2 -0.4);
						
						\fill[black] (-10pt + 0.0cm, -10pt - 5.5cm) rectangle (10pt +0.0cm, 10pt - 5.5cm);
						\node[text=white] at (0, -5.5) {$8$};
						\fill[black] (-10pt + 1.5cm, -10pt - 5.5cm) rectangle (10pt +1.5cm, 10pt - 5.5cm);
						\node[text=white] at (1.5, -5.5) {$6$};
						\fill[black] (-10pt + 3.0cm, -10pt - 5.5cm) rectangle (10pt +3.0cm, 10pt - 5.5cm);
						\node[text=white] at (3.0, -5.5) {$4$};
						\fill[black] (-10pt + 4.5cm, -10pt - 5.5cm) rectangle (10pt +4.5cm, 10pt - 5.5cm);
						\node[text=white] at (4.5, -5.5) {$2$};
						\fill[black] (-10pt + 6.0cm, -10pt - 5.5cm) rectangle (10pt +6.0cm, 10pt - 5.5cm);
						\node[text=white] at (6.0, -5.5) {$0$};
						
						\draw[->, thick] (0 + 0.1, -5.5 + 0.5) -- (0 + 0.75 -0.15, -2 -0.6);
						\draw[->, thick] (3 + 0.1, -5.5 + 0.5) -- (3 + 0.75 -0.15, -2 -0.6);
						\draw[->, thick] (6 + 0.1, -5.5 + 0.5) -- (6 + 0.75 -0.15, -2 -0.6);
						\draw[->, thick] (1.5 - 0.1, -5.5 + 0.5) -- (0 + 0.75 +0.15, -2 -0.6);
						\draw[->, thick] (4.6 - 0.1, -5.5 + 0.5) -- (3 + 0.75 +0.15, -2 -0.6);
						\draw[->, thick] (1.5 + 0.1, -5.5 + 0.5) -- (1.5 + 0.75 - 0.1, -3.5 -0.5);
						\draw[->, thick] (4.5 + 0.1, -5.5 + 0.5) -- (4.5 + 0.75 - 0.1, -3.5 -0.5);
						\draw[->, thick] (3.0 - 0.1, -5.5 + 0.5) -- (1.5 + 0.75 + 0.1, -3.5 -0.5);
						\draw[->, thick] (6.0 - 0.1, -5.5 + 0.5) -- (4.5 + 0.75 + 0.1, -3.5 -0.5);
						
						
						\draw (0 + 0.75, -7.5) circle[radius=10pt];
						\node at (0 + 0.75, -7.5) {$7$};
						\draw (1.5 + 0.75, -7.5) circle[radius=10pt];
						\node at (1.5 + 0.75, -7.5) {$5$};
						\draw (3.0 + 0.75, -7.5) circle[radius=10pt];
						\node at (3.0 + 0.75, -7.5) {$3$};
						\draw (4.5 + 0.75, -7.5) circle[radius=10pt];
						\node at (4.5 + 0.75, -7.5) {$1$};
						
						\draw[->, thick] (0 + 0.75 - 0.1, -7.5 + 0.5) -- (0 + 0.1, -5.5 - 0.5);
						\draw[->, thick] (0 + 0.75 + 0.1, -7.5 + 0.5) -- (1.5 - 0.1, -5.5 - 0.5);
						\draw[->, thick] (1.5 + 0.75 - 0.1, -7.5 + 0.5) -- (1.5 + 0.1, -5.5 - 0.5);
						\draw[->, thick] (1.5 + 0.75 + 0.1, -7.5 + 0.5) -- (3.0 - 0.1, -5.5 - 0.5);
						\draw[->, thick] (3.0 + 0.75 - 0.1, -7.5 + 0.5) -- (3.0 + 0.1, -5.5 - 0.5);
						\draw[->, thick] (3.0 + 0.75 + 0.1, -7.5 + 0.5) -- (4.5 - 0.1, -5.5 - 0.5);
						\draw[->, thick] (4.5 + 0.75 - 0.1, -7.5 + 0.5) -- (4.5 + 0.1, -5.5 - 0.5);
						\draw[->, thick] (4.5 + 0.75 + 0.1, -7.5 + 0.5) -- (6.0 - 0.1, -5.5 - 0.5);
						
						\fill[black] (0,-9.5) circle[radius=10pt];
						\node[text=white] at (0,-9.5) {$8$};
						\fill[black] (1.5,-11.0) circle[radius=10pt];
						\node[text=white] at (1.5,-11.0) {$6$};
						\fill[black] (3,-9.5) circle[radius=10pt];
						\node[text=white] at (3,-9.5) {$4$};
						\fill[black] (4.5,-11.0) circle[radius=10pt];
						\node[text=white] at (4.5,-11.0) {$2$};
						\fill[black] (6,-9.5) circle[radius=10pt];
						\node[text=white] at (6,-9.5) {$0$};
						
						\draw[->, thick] (0 + 0.1, -9.5 + 0.5) -- (0 + 0.75 - 0.1, -7.5 -0.5);
						\draw[->, thick] (3 + 0.1, -9.5 + 0.5) -- (3 + 0.75 - 0.1, -7.5 -0.5);
						\draw[->, thick] (3 - 0.1, -9.5 + 0.5) -- (1.5 + 0.75 + 0.1, -7.5 -0.5);
						\draw[->, thick] (6 - 0.1, -9.5 + 0.5) -- (4.5 + 0.75 + 0.1, -7.5 -0.5);
						\draw[->, dashed, thick] (1.5 -0.4, -11 + 0.3) -- (0 + 0.3, -9.5 -0.4);
						\draw[->, dashed, thick] (1.5 +0.4, -11 + 0.3) -- (3 - 0.3, -9.5 -0.4);
						\draw[->, dashed, thick] (1.5+3 -0.4, -11 + 0.3) -- (3 + 0.3, -9.5 -0.4);
						\draw[->, dashed, thick] (1.5+3 +0.4, -11 + 0.3) -- (6 - 0.3, -9.5 -0.4);
						
						\draw[black] (-10pt + 0.0cm + 0.75cm, -10pt - 13cm) rectangle (10pt +0.0cm +0.75cm, 10pt - 13cm);
						\node[text=black] at (0 + 0.75, -13) {$7$};
						\draw[black] (-10pt + 1.5cm +0.75cm, -10pt - 13cm) rectangle (10pt +1.5cm + 0.75cm, 10pt - 13cm);
						\node[text=black] at (1.5 + 0.75, -13) {$5$};
						\draw[black] (-10pt + 3.0cm + 0.75cm, -10pt - 13cm) rectangle (10pt +3.0cm + 0.75cm, 10pt - 13cm);
						\node[text=black] at (3.0 + 0.75, -13) {$3$};
						\draw[black] (-10pt + 4.5cm + 0.75cm, -10pt - 13cm) rectangle (10pt +4.5cm + 0.75cm, 10pt - 13cm);
						\node[text=black] at (4.5 + 0.75, -13) {$1$};
						
						\draw[->, thick] (3 - 0.75 + 0.1, -13 + 0.5) -- (3 -0.15, -9.5 -0.6);
						\draw[->, thick] (6 - 0.75 + 0.1, -13 + 0.5) -- (6 -0.15, -9.5 -0.6);
						\draw[->, thick] (1.5 - 0.75 - 0.1, -13 + 0.5) -- (0 +0.15, -9.5 -0.6);
						\draw[->, thick] (4.6 - 0.75 - 0.1, -13 + 0.5) -- (3 +0.15, -9.5 -0.6);
						\draw[->, thick] (1.5 - 0.75 + 0.1, -13 + 0.5) -- (1.5 - 0.1, -11 -0.5);
						\draw[->, thick] (4.5 - 0.75 + 0.1, -13 + 0.5) -- (4.5- 0.1, -11 -0.5);
						\draw[->, thick] (3.0 - 0.75 - 0.1, -13 + 0.5) -- (1.5 + 0.1, -11 -0.5);
						\draw[->, thick] (6.0 - 0.75 - 0.1, -13 + 0.5) -- (4.5 + 0.1, -11 -0.5);
						
						\draw[->, thick] (8.0 + 1.5, 0.5) -- (8.0 + 1.5, 1) -- (6.0 + 0.5 + 1.5, 1) -- (6.0 + 0.5 + 1.5, -5.5) -- (6.0 + 0.5, -5.5);
						\draw[thick] (8.0 + 3.0, 0.5) -- (8.0 + 3.0, 1) -- (8.0 + 1.5, 1);
						\draw[thick] (8.0 + 4.5, 0.5) -- (8.0 + 4.5, 1) -- (8.0 + 3.0, 1);
						\draw[thick] (8.0 + 6.0, 0.5) -- (8.0 + 6.0, 1) -- (8.0 + 4.5, 1);
						
						\draw [decorate, decoration={brace, amplitude=5pt}] (-0.75 , 1.0) -- (6 + 0.75, 1.0);
						\node[align=center, text width=3cm] at (3cm, 2cm) {\small \textbf{Circle Section}};
						
						\draw [decorate, decoration={brace, amplitude=5pt}] (-0.75 +9.5 , 1.5) -- (15.5, 1.5);
						\node[align=center, text width=3cm] at (12.125cm, 2.5cm) {\small \textbf{Radial Section}};

						\fill[black] (0 + 9.5,0) circle[radius=10pt];
						\node[text=white] at (+ 9.5,0) {$0$};
						\fill[black] (1.5+ 9.5,0) circle[radius=10pt];
						\node[text=white] at (1.5+ 9.5,0) {$2$};
						\fill[black] (3+ 9.5,0) circle[radius=10pt];
						\node[text=white] at (3+ 9.5,0) {$4$};
						\fill[black] (4.5+ 9.5,0) circle[radius=10pt];
						\node[text=white] at (4.5+ 9.5,0) {$6$};
						
						\draw (0 + 0.75+ 9.5, -2) circle[radius=10pt];
						\node at (0 + 0.75+ 9.5, -2) {$1$};
						\draw (3 + 0.75+ 9.5, -2) circle[radius=10pt];
						\node at (3 + 0.75+ 9.5, -2) {$5$};
						
						\draw (1.5 + 0.75+ 9.5, -3.5) circle[radius=10pt];
						\node at (1.5 + 0.75+ 9.5, -3.5) {$3$};
						\draw (4.5 + 0.75+ 9.5, -3.5) circle[radius=10pt];
						\node at (4.5 + 0.75+ 9.5, -3.5) {$7$};
						
						\draw[->, thick] (0.75 - 0.1 + 9.5, -1.5) -- (0 + 0.0+ 9.5, -0.5);
						\draw[->, thick] (0.75 + 0.1+ 9.5, -1.5) -- (1.5 - 0.1+ 9.5, -0.5);
						\draw[->, thick] (0.75+3 - 0.1+ 9.5, -1.5) -- (3 + 0.1+ 9.5, -0.5);
						\draw[->, thick] (0.75+3 + 0.1+ 9.5, -1.5) -- (4.5 - 0.1+ 9.5, -0.5);
						
						\draw[->, dashed, thick] (1.5 + 0.75 -0.4 + 9.5, -3.5 + 0.3) -- (0 + 0.75 + 0.3+ 9.5, -2 -0.4);
						\draw[->, dashed, thick] (1.5 + 0.75 +0.4+ 9.5, -3.5 + 0.3) -- (3 + 0.75 - 0.3+ 9.5, -2 -0.4);
						\draw[->, dashed, thick] (1.5+3 + 0.75 -0.4+ 9.5, -3.5 + 0.3) -- (3 + 0.75 + 0.3+ 9.5, -2 -0.4);
						\draw[->, dashed, thick] (1.5+3 + 0.75 -0.5+ 9.5, -3.5 + 0.1) -- (0 + 0.75 + 0.3+ 9.5 + 0.3, -2 -0.3);

						\fill[black] (-10pt + 0.0cm+ 9.5cm, -10pt - 5.5cm) rectangle (10pt +0.0cm+ 9.5cm, 10pt - 5.5cm);
						\node[text=white] at (0 + 9.5, -5.5) {$0$};
						\fill[black] (-10pt + 1.5cm+ 9.5cm, -10pt - 5.5cm) rectangle (10pt +1.5cm+ 9.5cm, 10pt - 5.5cm);
						\node[text=white] at (1.5+ 9.5, -5.5) {$2$};
						\fill[black] (-10pt + 3.0cm+ 9.5cm, -10pt - 5.5cm) rectangle (10pt +3.0cm+ 9.5cm, 10pt - 5.5cm);
						\node[text=white] at (3.0+ 9.5, -5.5) {$4$};
						\fill[black] (-10pt + 4.5cm + 9.5cm, -10pt - 5.5cm) rectangle (10pt +4.5cm+ 9.5cm, 10pt - 5.5cm);
						\node[text=white] at (4.5 + 9.5, -5.5) {$6$};
						
						\draw[->, thick] (0 - 0.1 + 9.5, -5.5 + 0.5) -- (0 + 0.75 -0.15 + 9.5, -2 -0.6);
						\draw[->, thick] (3 + 0.1 + 9.5, -5.5 + 0.5) -- (3 + 0.75 -0.15 + 9.5, -2 -0.6);
						\draw[->, thick] (1.5 - 0.1 + 9.5, -5.5 + 0.5) -- (0 + 0.75 +0.15 + 9.5, -2 -0.6);
						\draw[->, thick] (4.6 - 0.1 + 9.5, -5.5 + 0.5) -- (3 + 0.75 +0.15 + 9.5, -2 -0.6);
						\draw[->, thick] (1.5 + 0.1 + 9.5, -5.5 + 0.5) -- (1.5 + 0.75 - 0.1 + 9.5, -3.5 -0.5);
						\draw[->, thick] (4.5 + 0.1 + 9.5, -5.5 + 0.5) -- (4.5 + 0.75 - 0.1 + 9.5, -3.5 -0.5);
						\draw[->, thick] (3.0 - 0.1 + 9.5, -5.5 + 0.5) -- (1.5 + 0.75 + 0.1 + 9.5, -3.5 -0.5);
						\draw[->, thick] (0 + 0.1 + 9.5, -5.5 + 0.5) -- (4.5 + 0.75 - 0.4 + 9.5, -3.5 -0.3);

						\draw (0 + 0.75 +9.5, -7.5) circle[radius=10pt];
						\node at (0 + 0.75+9.5, -7.5) {$1$};
						\draw (1.5 + 0.75+9.5, -7.5) circle[radius=10pt];
						\node at (1.5 + 0.75+9.5, -7.5) {$3$};
						\draw (3.0 + 0.75+9.5, -7.5) circle[radius=10pt];
						\node at (3.0 + 0.75+9.5, -7.5) {$5$};
						\draw (4.5 + 0.75+9.5, -7.5) circle[radius=10pt];
						\node at (4.5 + 0.75+9.5, -7.5) {$7$};
						
						\draw[->, thick] (0 + 0.75 - 0.1+9.5, -7.5 + 0.5) -- (0 - 0.1+9.5, -5.5 - 0.5);
						\draw[->, thick] (0 + 0.75 + 0.1+9.5, -7.5 + 0.5) -- (1.5 - 0.1+9.5, -5.5 - 0.5);
						\draw[->, thick] (1.5 + 0.75 - 0.1+9.5, -7.5 + 0.5) -- (1.5 + 0.1+9.5, -5.5 - 0.5);
						\draw[->, thick] (1.5 + 0.75 + 0.1+9.5, -7.5 + 0.5) -- (3.0 - 0.1+9.5, -5.5 - 0.5);
						\draw[->, thick] (3.0 + 0.75 - 0.1+9.5, -7.5 + 0.5) -- (3.0 + 0.1+9.5, -5.5 - 0.5);
						\draw[->, thick] (3.0 + 0.75 + 0.1+9.5, -7.5 + 0.5) -- (4.5 - 0.1+9.5, -5.5 - 0.5);
						\draw[->, thick] (4.5 + 0.75 + 0.1+9.5, -7.5 + 0.5) -- (4.5 + 0.1+9.5, -5.5 - 0.5);
						\draw[->, thick] (4.5 + 0.75 - 0.4 + 9.5, -7.5 + 0.4) -- (0 + 0.4+9.5, -5.5 - 0.5);

						\fill[black] (0 +9.5,-9.5) circle[radius=10pt];
						\node[text=white] at (0+9.5,-9.5) {$0$};
						\fill[black] (1.5+9.5,-11.0) circle[radius=10pt];
						\node[text=white] at (1.5+9.5,-11.0) {$2$};
						\fill[black] (3+9.5,-9.5) circle[radius=10pt];
						\node[text=white] at (3+9.5,-9.5) {$4$};
						\fill[black] (4.5+9.5,-11.0) circle[radius=10pt];
						\node[text=white] at (4.5+9.5,-11.0) {$6$};
						
						\draw[->, thick] (0 - 0.1 + 9.5, -9.5 + 0.5) -- (0 + 0.75 - 0.1+ 9.5, -7.5 -0.5);
						\draw[->, thick] (3 + 0.1+ 9.5, -9.5 + 0.5) -- (3 + 0.75 - 0.1+ 9.5, -7.5 -0.5);
						\draw[->, thick] (3 - 0.1+ 9.5, -9.5 + 0.5) -- (1.5 + 0.75 + 0.1+ 9.5, -7.5 -0.5);
						\draw[->, thick] (0 + 0.4 + 9.5, -9.5 + 0.2) -- (4.5 + 0.75 - 0.3+ 9.5, -7.5 -0.4);
						\draw[->, dashed, thick] (1.5 -0.4+ 9.5, -11 + 0.3) -- (0 + 0.3+ 9.5, -9.5 -0.4);
						\draw[->, dashed, thick] (1.5 +0.4+ 9.5, -11 + 0.3) -- (3 - 0.3+ 9.5, -9.5 -0.4);
						\draw[->, dashed, thick] (1.5+3 -0.4+ 9.5 + 0.2, -11 + 0.45) -- (3 + 0.3+ 9.5, -9.5 -0.4);
						\draw[->, dashed, thick] (1.5+3 -0.4+ 9.5, -11 + 0.3) -- (0 + 0.5+ 9.5 +0.1, -9.5 -0.25);

						\draw[black] (-10pt + 0.0cm + 0.75cm + 9.5cm, -10pt - 13cm) rectangle (10pt +0.0cm +0.75cm + 9.5cm, 10pt - 13cm);
						\node[text=black] at (0 + 0.75 + 9.5, -13) {$1$};
						\draw[black] (-10pt + 1.5cm +0.75cm + 9.5cm, -10pt - 13cm) rectangle (10pt +1.5cm + 0.75cm + 9.5cm, 10pt - 13cm);
						\node[text=black] at (1.5 + 0.75 + 9.5, -13) {$3$};
						\draw[black] (-10pt + 3.0cm + 0.75cm + 9.5cm, -10pt - 13cm) rectangle (10pt +3.0cm + 0.75cm + 9.5cm, 10pt - 13cm);
						\node[text=black] at (3.0 + 0.75 + 9.5, -13) {$5$};
						\draw[black] (-10pt + 4.5cm + 0.75cm + 9.5cm, -10pt - 13cm) rectangle (10pt +4.5cm + 0.75cm + 9.5cm, 10pt - 13cm);
						\node[text=black] at (4.5 + 0.75 + 9.5, -13) {$7$};

						\draw[->, thick] (3 - 0.75 + 0.1 + 9.5, -13 + 0.5) -- (3 -0.15 + 9.5, -9.5 -0.6);
						\draw[->, thick] (1.5 - 0.75 - 0.1 + 9.5, -13 + 0.5) -- (0 +0.15 + 9.5, -9.5 -0.6);
						\draw[->, thick] (4.6 - 0.75 - 0.1 + 9.5, -13 + 0.5) -- (3 +0.15 + 9.5, -9.5 -0.6);
						\draw[->, thick] (1.5 - 0.75 + 0.1 + 9.5, -13 + 0.5) -- (1.5 - 0.1 + 9.5, -11 -0.5);
						\draw[->, thick] (4.5 - 0.75 + 0.1 + 9.5, -13 + 0.5) -- (4.5- 0.1 + 9.5, -11 -0.5);
						\draw[->, thick] (3.0 - 0.75 - 0.1 + 9.5, -13 + 0.5) -- (1.5 + 0.1 + 9.5, -11 -0.5);
						\draw[->, thick] (6.0 - 0.75 + 0.1 + 9.5, -13 + 0.5) -- (4.5 + 0.1 + 9.5, -11 -0.5);
						
						\draw[->, thick] (6 - 0.75 - 0.1 + 9.5, -13 + 0.5) -- (0 +0.15 + 9.5 + 0.4, -9.5 -0.5);
						
					\end{scope}
				\end{tikzpicture}
			\end{adjustwidth}
			\caption{\textbf{Parallelization of the smoother using the Give approach.} 
				The figure represents a grid which consists of nine circular and eight radial smoother. 
				In this example, inner- and outermost circle smoothed with a circle pattern are colored black. 
				Vertices represent computational tasks corresponding to a line of nodes. 
				Solid edges visualize dependencies between tasks as given by Eq.~\eqref{eq:relaxation}.
				Dashed edges represent synthetic dependencies introduced to avoid concurrent updates of the same memory location and ensure conflict-free execution during parallel processing.
			}
			\label{fig:smoother_parallel}
		\end{figure}
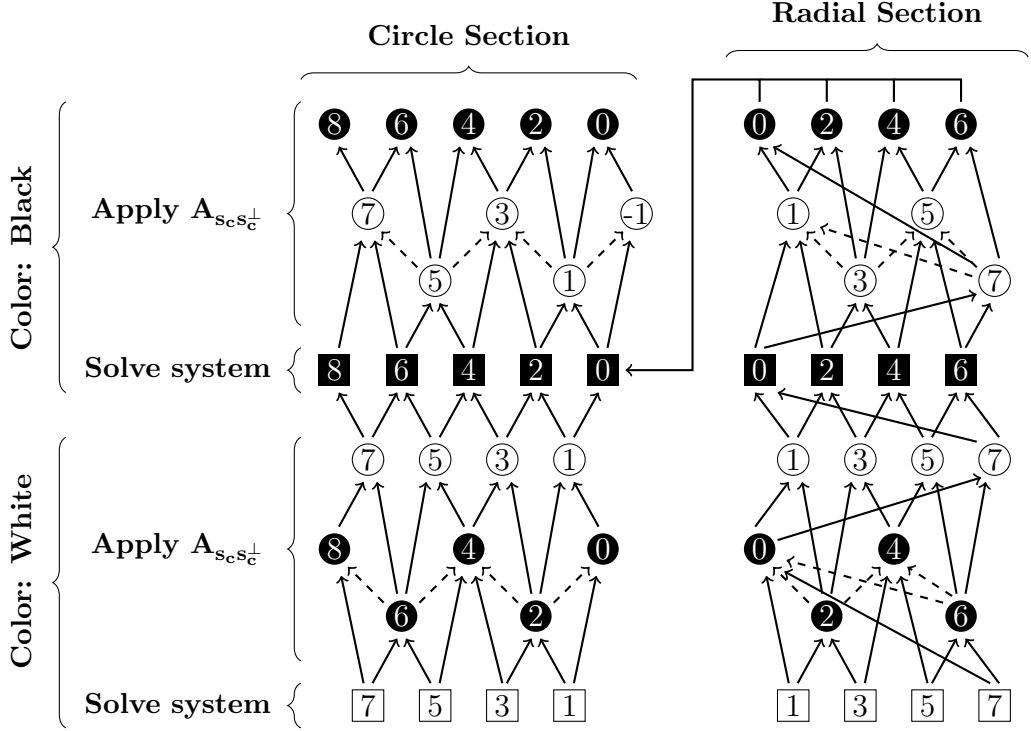

		For the smoothing operation on multigrid level $i\in\{0,\ldots,L-1\}$, we solve subsystems of the original system with matrix $A^{(i)}$, right hand side $f^{(i)}$, and solution $u^{(i)}$. 
		These systems write
		\begin{align}
			\label{eq:relaxation}
			A^{(i)}_{s_cs_c}u^{(i)}_{s_c} =  f^{(i)}_{s_c} - A^{(i)}_{s_cs_c^{\perp}}u^{(i)}_{s_c^{\perp}} ,
		\end{align}
		where $s$ refers to the smoothing operation, either circle or radial, and $c$ to the color, either black or white. 
		Furthermore, $A^{(i)}_{s_cs_c}$, $u^{(i)}_{s_c}$, and $f^{(i)}_{s_c}$ correspond to the nodes to be smoothed and the complementary part $A^{(i)}_{s_cs_c^{\perp}}$ and $u^{(i)}_{s_c^{\perp}}$ to the nodes connected, which contribute to the right hand side of the system. 
		In the Take approach, the complementary updates to the right hand side can also be done in parallel. 
		While lines of the same smoother and color can be solved completely in parallel, several dependencies to update the right hand sides have to be considered with the Give approach. 
		Using the Give approach, we obtain a more complex parallelization pattern as nodes change the values of their neighbors. 
		In the prior (Give) implementation, the additional dependencies to be added to~\cref{fig:applyA_parallelization} can be found in~\cite[Fig. 8]{LSK25}. 
		The novel implementation adopts a slightly less complex pattern as visualized in~\cref{fig:smoother_parallel}. 
		In this approach, the rows for the application of the complementary part are executed in parallel pattern of 2-4-4 with barriers in-between, meaning that, first, every second line is treated, then, two sweeps of lines with a distance of four are executed before, eventually, the system is solved for every second line in parallel.
		
		\subsection{Multigrid features}\label{sec:newmg}
		
		GMGPolar now also supports $W$- and $F$-cycles. 
		The $W$-cycle performs additional coarse grid corrections by revisiting intermediate levels during the upward traversal, yielding a more thorough error reduction on coarser grids. 
		With the $F$-cycle, GMGPolar offers a hybrid between the $V$-cycle and $W$-cycle. 
		It combines the efficiency of the $V$-cycle with the robustness of the $W$-cycle by selectively applying additional coarse-grid corrections; see~\cite{TOS01, Hackbusch}.
		
		In addition, GMGPolar now provides support for a full multigrid cycle (FMG). 
		FMG uses nested iterations to compute a refined initial approximation before standard multigrid cycles are applied. 
		With this approach, we significantly accelerate the convergence process, as shown in~\cref{sec:num}.

		\section{Numerical results}\label{sec:num}
		
\begin{figure}
			\begin{minipage}[t]{0.32\textwidth}
				\centering
				\includegraphics[width=\textwidth]{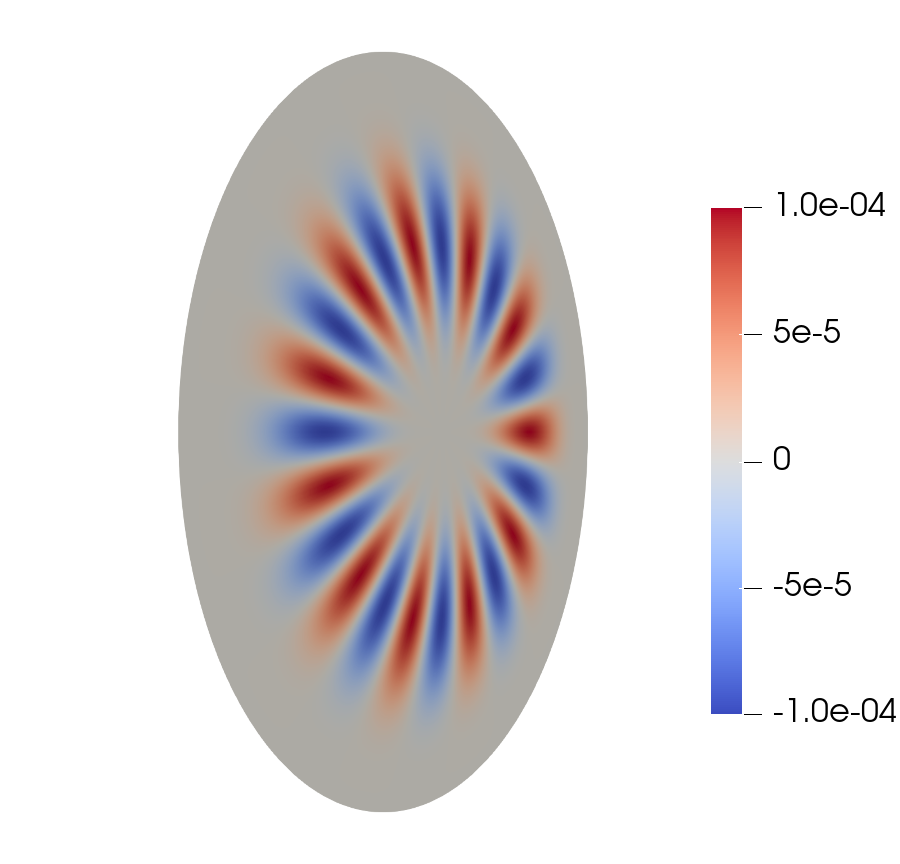}
			\end{minipage}
			\begin{minipage}[t]{0.32\textwidth}
				\centering
				\includegraphics[width=\textwidth]{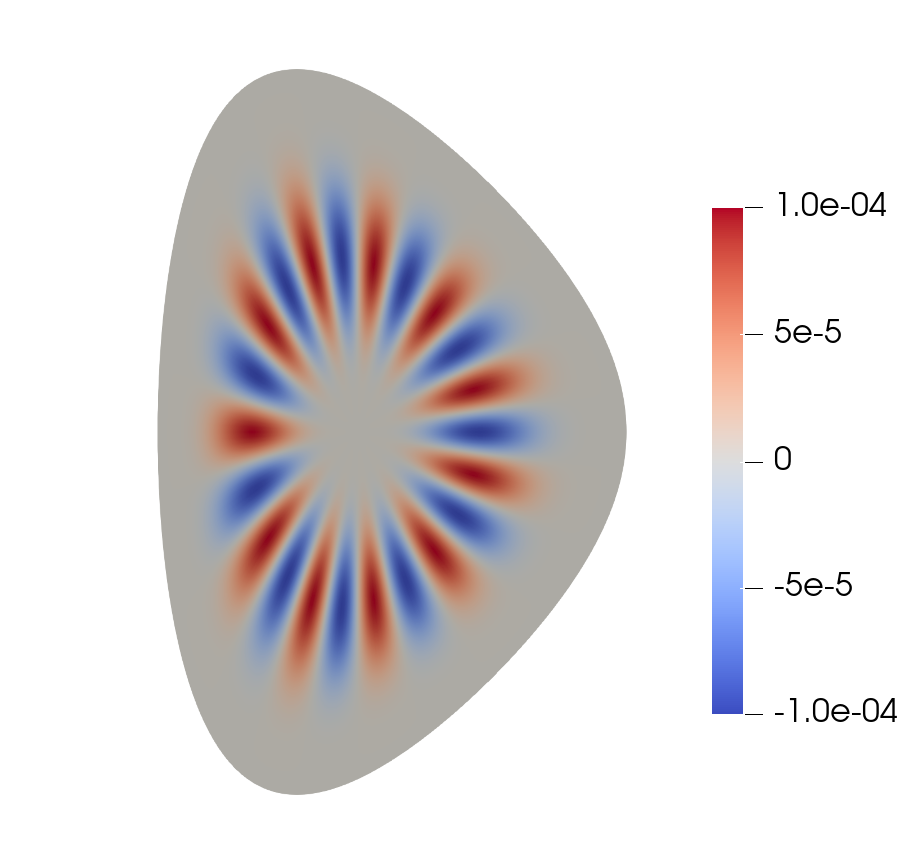}
			\end{minipage}
			\begin{minipage}[t]{0.32\textwidth}
				\centering
				\includegraphics[width=\textwidth]{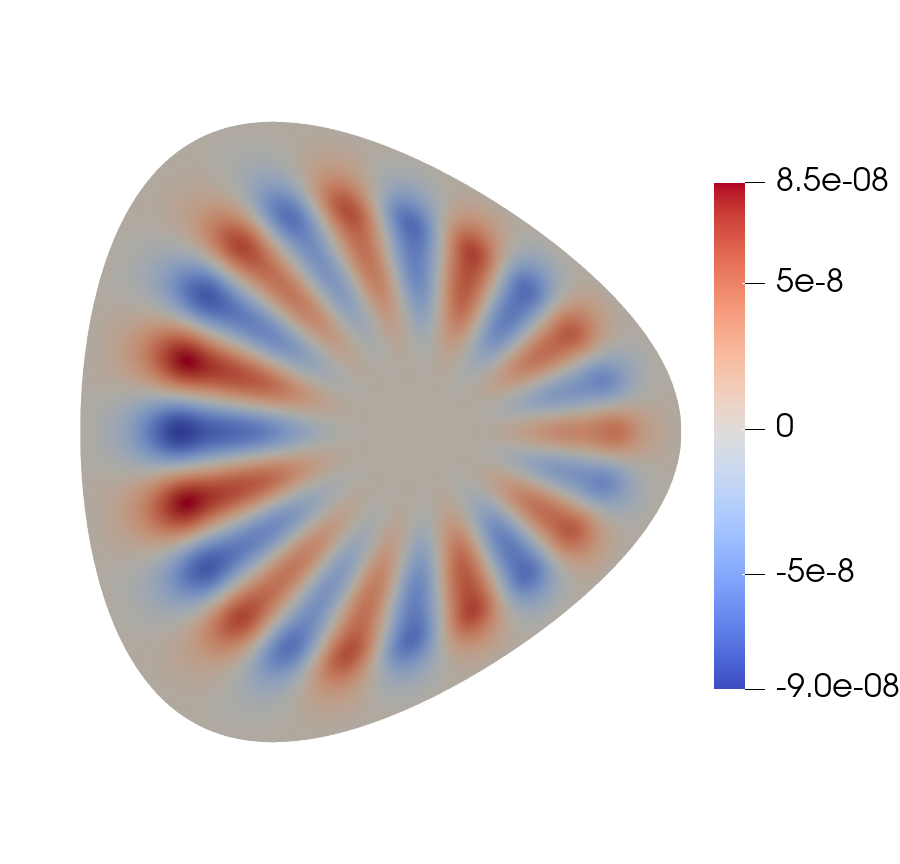}
			\end{minipage}
			\caption{\textbf{Illustration of the manufactured solution~\eqref{eq:sol_polar6}.} 
				Visualized solutions for the Shafranov (left), Czarny (center), and Culham (right) geometries.}
			\label{fig:sol_polar6}
		\end{figure}

		In this section, we will present numerical results for the model problem~\eqref{eq:poisson}. 
		The coefficients $\alpha$ and $\beta$ as well as the manufactured solution are inspired by the simulation of plasma in tokamak fusion reactors and taken from prior benchmarks in~\cite{BLK23,LSK25}. 
		We consider a \textit{Polar solution} with oscillations aligned with the polar grid
		\begin{align}\label{eq:sol_polar6}
			u(x, y) = 0.4096 \left(\frac{r(x, y)}{R_{max}} \right)^6 \left( 1 - \frac{r(x, y)}{R_{max}} \right)^6 \cos(11\vartheta(x,y)).
		\end{align} 
		\cref{fig:sol_polar6} illustrates the solution for the Shafranov (left), Czarny (center), and Culham (right) geometries. 
		Note that for the nonanalytical Culham geometry, no exact, manufactured solution is supplied. For the coefficients $\alpha$ and $\beta$, we set
		\begin{align}
			\alpha(r) = \exp\left[-\tanh \left( \frac{\frac{r(x, y)}{R_{max}} - r_p}{\delta_r} \right) \right],\quad
			{\beta(r)} = \frac{1}{\alpha(r)}, \label{eq:alpha_zoni}
		\end{align}
		where $\delta_r = 0.05$ and $r_p = 0.7$, as in~\cite{BLK23,LSK25}, and $R_{max}=1.3$ as in~\cite{KKR22,LSK25}.
		
		Our experiments were run on an AMD EPYC 7601, 2.2 GHz, node with two 32-Core sockets and 128 GB DDR4 RAM of the supercomputer CARA at the German Aerospace Center as well as on Intel Xeon "Skylake" Gold 6132, 2.60 GHz, node with four 14-Core sockets with 384GB DDR4 RAM of a small internal cluster. 
		As for coarse level solver, we use MUMPS v5.5.1~\cite{MUMPS}. 
		We use LIKWID~\cite{LIKWID,likwid_code} for measuring performance in MFLOPs/s and data transfer in MBytes/s.
		
		We allow a grid-adapted maximum number of levels yielding six multigrid levels for a grid of size ${193\times256}$ and 11 levels for a grid of size ${6\,145\times8\,192}$. 
		As in prior publications~\cite{KKR22,LSK25}, we use an anisotropic grid refinement, approximately where the gradient of coefficient $\alpha$ attains its minimum;~\cite[Fig.~1]{KKR22} or, symbolically, in~\cref{fig:geom}. 
		This setup demonstrates that GMGPolar is also capable of handling anisotropic grids efficiently. 
		The absolute or relative convergence criteria were selected depending on the purposes of the individual, following subsections. 
		Convergence is measured in the weighted $\|\cdot\|_{\ell_2}$-norm
		\begin{align*}
			\|v\|_{\ell_2} = \sqrt{{\frac{1}{n}}\sum_{i=1}^n v_i^2}.
		\end{align*}
		
		In \cref{sec:roofline}, we provide roofline model results. 
		In \cref{sec:nummemory}, we consider the memory requirements. 
		In \cref{sec:weak}, we provide weak scaling results. 
		In \cref{sec:strong}, we show strong scaling results. 
		In \cref{sec:cycles}, we provide results on FMG and different cycle types to obtain algorithmic speedups. \MK{Eventually, in~\cref{sec:pcg}, we provide some experimental results when using GMGPolar as a preconditioner for conjugated gradients.}
		
		\subsection{Roofline model}\label{sec:roofline}
		
		\begin{figure}
			\begin{center}
				\includegraphics[width=0.6\textwidth]{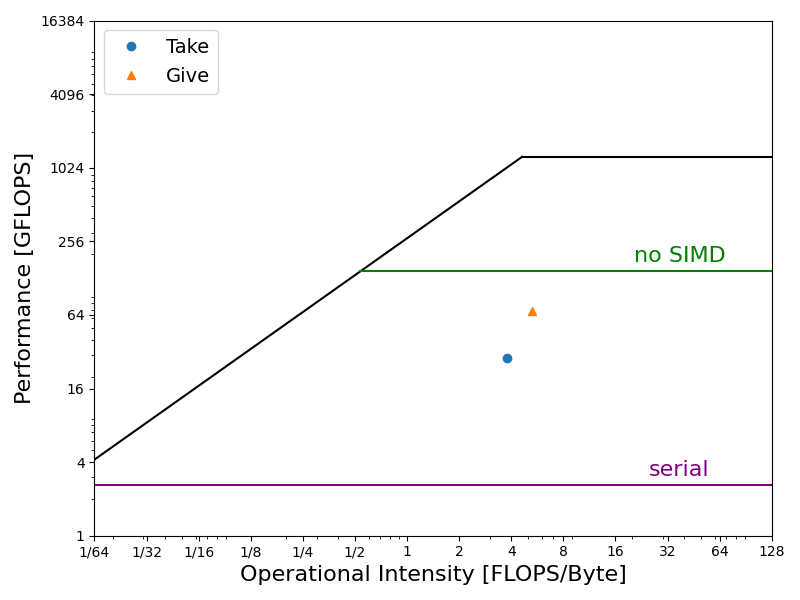}
			\end{center}
			\caption{\textbf{Roofline model of GMGPolar.} \textbf{Czarny} geometry with \textbf{Take} and \textbf{Give} stencil implementation and problem size $\mathbf{6145 \times 8192}$.}
			\label{fig:roofline}
		\end{figure}
		
		In this first subsection, we provide a roofline model for GMGPolar. Our roofline model uses the benchmarked peak performance against the operational intensity. This model shows the hardware limitations and the potential benefit in optimizing computational or memory aspects of the considered implementations. We use LIKWID~\cite{LIKWID,likwid_code} for measuring double precision computations in MFLOPs/s, the memory bandwidth in MBytes/s as well as the operational intensity in FLOPs/Byte. 
		We use the roofline model computed on a full node (56 cores) of an Intel Xeon "Skylake" Gold 6132, 2.60 GHz, with four 14-Core sockets. 
		The peak flops of 1256.6 GFlops/s were quantified with the LIKWID benchmark \textit{peakflops\_avx} with settings \textit{N:1792kB:56}. 
		The value of 1792 kB was obtained from 32kB for each of the 56 hardware threads so that each vector chunk fits into the L1 cache of one core. 
		The maximum memory bandwith of 272 GBytes/s was obtained with the LIKWID benchmark \textit{stream\_mem\_avx\_fma} with settings \textit{N:2GB:56}. 
		As a test case, we consider the Czarny geometry with a resolution of $6145 \times 8192$ nodes. 
		As a convergence criterion, we choose a relative reduction of the initial residual by $1e8$.
		
		As indicated by our first findings in~\cite{BLK23,LSK25} and although being a sparse matrix-free implementation, we see that GMGPolar has a rather elevated computational intensity and is not memory bound; cf.~\cref{fig:roofline}. 
		Aside from the potential through AVX SIMD operations which is not yet exploited, we see that in particular the Give approach comes close to the compute limit. 
		We thus see that the matrix-free implementation benefits from storing several expensive function evaluations on the transformed geometry.
		
		\subsection{Memory requirements}\label{sec:nummemory}
		
		\begin{table}[b!]
			\footnotesize
			\hspace*{-0.8cm}
			\begin{tabular}[c]{ccccccc}
				\bottomrule
				$n_r\times n_\vartheta$ & 193 $\times$ 256 & 385$\times$512 & 769 $\times$ 1\,024 & 1\,537 $\times$ 2\,048 &  3\,073 $\times$ 4\,096 & 6\,145 $\times$ 8\,192 \\
				\toprule            
				\midrule            
				\multicolumn{7}{c}{\textbf{GMGPolar v1}}\\
				\midrule 
				\bottomrule            
				\textbf{Estimate} & \multirow{2}{*}{4.74 MB} & \multirow{2}{*}{18.92 MB} & \multirow{2}{*}{75.60 MB} & \multirow{2}{*}{302.19 MB} & \multirow{2}{*}{1\,208.35 MB} & \multirow{2}{*}{4\,832.62 MB} \\
				(Min. cache) & & & & & & \\
				\textbf{Give} & \multirow{2}{*}{4.96 MB} & \multirow{2}{*}{18.86 MB} & \multirow{2}{*}{73.96 MB} & \multirow{2}{*}{294.97 MB} & \multirow{2}{*}{N/A*} & \multirow{2}{*}{N/A*} \\
				(Min. cache) & & & & & & \\
				\toprule            
				\midrule
				\multicolumn{7}{c}{\textbf{GMGPolar v2}}\\         
				\midrule
				\bottomrule
				\textbf{Estimate} & \multirow{2}{*}{2.64 MB} & \multirow{2}{*}{10.51 MB} & \multirow{2}{*}{42.00 MB} & \multirow{2}{*}{167.88 MB} & \multirow{2}{*}{671.31 MM} & \multirow{2}{*}{	2\,684.79 MB} \\
				(Min. cache) & & & & & & \\            
				\textbf{Give} & \multirow{2}{*}{3.22 MB} & \multirow{2}{*}{12.24 MB} & \multirow{2}{*}{47.56 MB} & \multirow{2}{*}{189.04 MB} & \multirow{2}{*}{754.03 MB} & \multirow{2}{*}{3\,015.47 MB} \\
				(Min. cache) & & & & & & \\
				\textbf{Give} & \multirow{2}{*}{3.22 MB} & \multirow{2}{*}{12.25 MB} & \multirow{2}{*}{47.58 MB} & \multirow{2}{*}{189.09 MB} & \multirow{2}{*}{754.13 MB} & \multirow{2}{*}{3\,015.65 MB}  \\
				(Coeff.) & & & & & & \\
				\textbf{Give} & \multirow{2}{*}{5.19 MB} & \multirow{2}{*}{20.08 MB} & \multirow{2}{*}{78.89 MB} & \multirow{2}{*}{314.19 MB} & \multirow{2}{*}{1\,254.33 MB} & \multirow{2}{*}{5\,016.03 MB}   \\
				(Geom.) & & & & & & \\            
				\textbf{Give} & \multirow{2}{*}{5.20 MB} & \multirow{2}{*}{20.09 MB} & \multirow{2}{*}{78.90 MB} & \multirow{2}{*}{314.21 MB} & \multirow{2}{*}{1\,254.38 MB} & \multirow{2}{*}{5\,016.12 MB} \\
				(Coeff. \& Geom.) & & & & & & \\            
				\textbf{Take} & \multirow{2}{*}{5.20 MB} & \multirow{2}{*}{20.09 MB} & \multirow{2}{*}{78.90 MB} & \multirow{2}{*}{314.21 MB} & \multirow{2}{*}{1\,254.38 MB} & \multirow{2}{*}{5\,016.12 MB} \\
				(Coeff. \& Geom.) & & & & & & \\
				
				\bottomrule            
			\end{tabular}
			\caption{\textbf{Memory requirements of different novel GMGPolar implementations with comparison to old implementation.} \textbf{Czarny} geometry from grid size $\mathbf{193\times256}$ to grid size $\mathbf{6\,145\times8\,192}$. $n_r$ provides the resolution in the first dimension, $n_\vartheta$ provides the resolution in the second dimension. 
				\textit{Min. chache} stands for the evaluations of sine and cosine functions in one dimension which are always cached. 
				\textit{Coeff.} stands for caching of $\alpha$ and $\beta$ evaluations from~\eqref{eq:poisson} and~\eqref{eq:alpha_zoni} and \textit{Geom.} stands for caching transformation coefficients from~\eqref{eq:squarejacobian}.\\ 
				* This run was canceled / not executed as it took too long with the massif memory tool.
			}
			\label{tab:memory}
		\end{table}
		
		In this section, we consider the memory requirement of GMGPolar, comparing the novel implementation against the previous one. 
		We compare the new Give implementation where either nothing (\textit{Min. cache}), profile coefficients (\textit{Coeff.}), geometry transformations (\textit{Geom.}), or profile coefficients and geometry transformations (\textit{Coeff. \& Geom.}) are cached and with the Take approach where coefficients and geometries transformations are both cached; see the end of~\cref{sec:cache}. We, furthermore, provide estimates on the expected memory requirements as laid out in~\cref{sec:cache}.
		For these experiments, the direct solver MUMPS has been replaced by a custom-made solver as it could otherwise not be measured with valgrind's tool massif~\cite{nethercote2007valgrind,nethercote2006building}. 
		Furthermore, we replaced the right hand side by a constant vector equal to one and only three iterations were conducted. 
		It has to be noted that the corresponding custom-made solver was not designed to replace MUMPS for a performance-oriented execution but only provides a fallback implementation. 
		With this fallback implementation, the measured memory of the Give approach increases from approximately five vectors (on the finest level) to be stored to 5.6 vectors and for the Take approach from nine vectors to be stored to 9.4 vectors. 
		An optimized version of the custom-build solver is already available with a new pull request of GMGPolar.
		
		From~\cref{tab:memory}, we see that the memory requirements of the Give approach were reduced by approximately 36~\%, when compared to the prior implementation. 
		On the other hand, the implementation of the Take approach in the novel version 2 comes close to the requirements of the prior Give implementation. 
		We also see that the caching of the density profile coefficients from~\eqref{eq:poisson} and~\eqref{eq:alpha_zoni} is almost negligible with respect to memory requirements while the caching of the domain geometry transformations $a^{rr}$, $a^{r \vartheta}$, $a^{\vartheta \vartheta}$, and $\det DF_g$ from~\eqref{eq:squarejacobian} leads to a relevant memory increase.
		
		\subsection{Weak scaling}\label{sec:weak}
		
		In this section, we provide weak scaling experiments going from a single core to 64 cores on CARA.
		We start with a geometry of 769$\times$1024 nodes and end with a grid of size 6145$\times$8192, effectively scaling from approximately 800\,000 to 50 million nodes. For the Give stencil implementation, we use the default setting of caching the profile coefficient values. For the Take implementation, profile coefficient and geometry values are cached.
		As a convergence criterion, we choose a relative reduction of the initial residual by $1e8$. From~\cref{tab:weakscaling}, we obtain weak scaling efficiencies of 25.64~\% and 41.06~\% for the Shafranov and Czarny geometry and the Give implementation. 
		The corresponding values were 29.71~\% and 46.14~\% for a very similar test case in the previous implementation; cf.~\cite{LSK25}. 
		However, with the algorithm itself substantially sped up by a factor of two to four (cf.~\cref{fig:strong_comparison_compute}), the results appear acceptable. 
		For the Culham geometry, we obtain a weak scaling efficiency of 72.95~\% from one to 64 cores. 
		From one to 16 cores, we obtain weak scaling efficiencies of 69.91 to 91.50~\% for the three different geometries. 
		For the Take implementation, we obtain worse weak scaling results but can substantially speed up the computation by factors of 2.7 to 14.6. 
		As indicated by the roofline model, we see that the Take approach, where more information is stored in memory, is the more advantageous the more complex the geometry is.
		
		\begin{table}[t!]
			\small
			\begin{center}
				\begin{tabular}[c]{ccccccccc}
					\toprule
					\multicolumn{8}{c}{\textbf{Shafranov geometry}}\\
					\bottomrule
					\midrule
					$n_r\times n_\vartheta$ & \textnormal{Cores} & \textnormal{Stencil} & \textnormal{Time} & \textnormal{Eff.} & \textnormal{Stencil} & \textnormal{Time} & \textnormal{Eff.} & \textnormal{G/T} \\
					\midrule
					\phantom{0\,}769 $\times$ 1\,024 & 1 &
					\multirow{4}{*}{Give} & 32.42 s  & 100~\%  & 
					\multirow{4}{*}{Take} & 11.83 s & 100~\% & 2.74 \\
					1\,537 $\times$ 2\,048 & 4  & & 34.67 s & 93.51~\% &  & 13.39 s & 88.35~\% & 2.58 \\
					3\,073 $\times$ 4\,096 & 16 & & 46.37 s & {69.91~\%} &  & {25.98 s} & 45.54~\% & 1.78 \\
					6\,145 $\times$ 8\,192 & 64 & & 126.41 s& 25.64~\% &  & 112.32 s & 10.53~\% & 1.12\\
					\bottomrule
					
					\midrule
					\multicolumn{8}{c}{\textbf{Czarny geometry}}\\
					\bottomrule
					\midrule
					$n_r\times n_\vartheta$ & \textnormal{Cores} & \textnormal{Stencil} & \textnormal{Time} & \textnormal{Eff.} & \textnormal{Stencil} & \textnormal{Time} & \textnormal{Eff.} & \textnormal{G/T} \\
					\midrule
					\phantom{0\,}769 $\times$ 1\,024 & 1 &
					\multirow{4}{*}{Give} & 41.77 s & 100~\%  & 
					\multirow{4}{*}{Take} & 9.18 s & 100~\% & 4.55 \\
					1\,537 $\times$ 2\,048 & 4  &  & 43.60 s & 95.80~\% &  & 10.16 s & 90.35~\% & 4.29 \\
					3\,073 $\times$ 4\,096 & 16 &  & 53.00 s & 78.81~\% &  & 19.48 s & 47.12~\% & 2.72 \\
					6\,145 $\times$ 8\,192 & 64 &  & 101.72 s & 41.06~\% &  & 77.89 s & 11.78~\% & 1.30\\
					\bottomrule
					
					\midrule
					\multicolumn{8}{c}{\textbf{Culham geometry}}\\
					\bottomrule
					\midrule
					$n_r\times n_\vartheta$ & \textnormal{Cores} & \textnormal{Stencil} & \textnormal{Time} & \textnormal{Eff.} & \textnormal{Stencil} & \textnormal{Time} & \textnormal{Eff.} & \textnormal{G/T}\\
					\midrule
					\phantom{0\,}769 $\times$ 1\,024 & 1 &
					\multirow{4}{*}{Give} &  115.52 s & 100~\%  & 
					\multirow{4}{*}{Take} & 7.90 s & 100~\% &  14.62 \\
					1\,537 $\times$ 2\,048 & 4  &  & 118.38 s & 97.58~\% &  & 8.64 s & 91.43~\% &  13.70\\
					3\,073 $\times$ 4\,096 & 16 &  & 126.26 s & 91.50~\% &  & 15.40 s & 51.29~\% &  8.19\\
					6\,145 $\times$ 8\,192 & 64 &  & 158.36 s & 72.95~\% &  & 63.29 s &  12.48~\% & 2.50\\
					\bottomrule            
				\end{tabular}
				\caption{\textbf{Weak scaling of GMGPolar for different geometries and the two different stencil implementations.}
					$n_r$ provides the resolution in the first dimension, $n_\vartheta$ provides the resolution in the second dimension, \textit{Cores} provides the numbers of cores used, \textit{Time} provides the total solver time, \textit{Eff.} provides the weak scaling efficiency computed with respect to the single core run on grid size $\mathbf{769 \times 1\,024}$ and 64 cores on $\mathbf{6\,145 \times 8\,192}$, and \textit{G/T} provides the speedup obtained by using the Take approach instead of the Give approach.}
				\label{tab:weakscaling}
			\end{center}
		\end{table}
		
		\subsection{Strong scaling}\label{sec:strong}
		
		In this section, we first present strong scaling results for the novel GMGPolar implementation on different geometries. 
		Eventually, we also provide a comparison with respect to scaling behavior of our prior implementation of~\cite{BLK23,LSK25}. 
		As a convergence criterion, we choose a relative reduction of the initial residual by $1e8$. 
		Our experiments were again run on CARA. We double the number of cores from one to 64 and consider the strong scaling behavior solver timings of GMGPolar v2 and, in comparison, setup and solver timings of the novel implementation against GMGPolar v1. 
		
\begin{figure}
			\begin{minipage}[t]{1\textwidth}
				\centering
				\includegraphics[trim={2cm 2cm 2cm 2cm},clip,width=\textwidth]{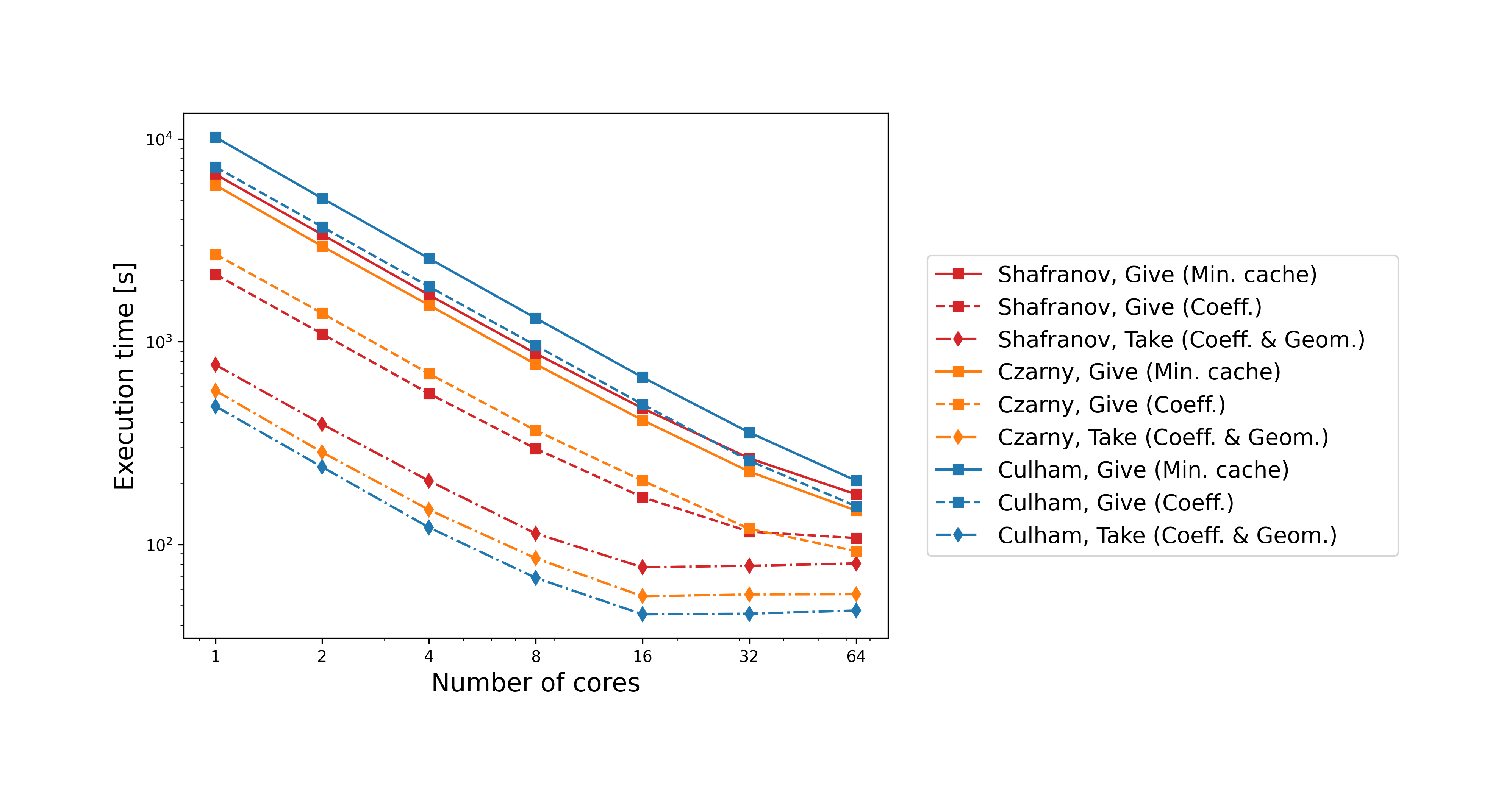}
			\end{minipage}
			\caption{\textbf{Strong scaling and runtime comparison of GMGPolar for different geometries and stencil and caching implementations.} 
				Visualization of strong scaling for \textbf{Shafranov}, \textbf{Czarny}, and \textbf{Culham} geometry with \textbf{Give} and \textbf{Take} stencil implementations on problem size $\mathbf{6145 \times 8192}$ with convergence criterion of a relative reduction of the initial residual by $1e8$.}
			\label{fig:strong_geometries}
		\end{figure}

		In~\cref{fig:strong_geometries}, we provide computing times and strong scaling of the novel GMGPolar on Shafranov, Czarny, and Culham geometry, respectively, with a grid size of 6145 $\times$ 8192, i.e., approximately 50 million degrees of freedom.
		We first see that additional caching of the profile coefficients in the Give approach, which only minimally increases the requirement memory (see~\cref{{tab:memory}}), substantially reduces the runtime of the Give approach. 
		Additionally, the Take approach, where coefficients and geometry information is stored, drastically reduces the runtime -- at the cost of approximately 66~\% of additional memory. 
		However, we also see that the different geometries benefit differently from storing coefficient and geometry information. Intuitively, the more complex the geometry, the less beneficial is the sole caching of the coefficients, recomputing geometry-dependent values. 
		For the faster Take implementation, we see that parallelization stagnates after 16 cores. This means that not enough data is available with 6145$\times$8192 nodes for the compute parallelism offered through 32 to 64 cores and that, at best, four of these cross sections could be computed on a single node; cf.~\cref{fig:tokamak_cross_270}.
		
		\begin{figure}[h!]
			\begin{minipage}[t]{0.48\textwidth}
				\centering
				\includegraphics[trim={2cm 2cm 2cm 2cm},clip,width=\textwidth]{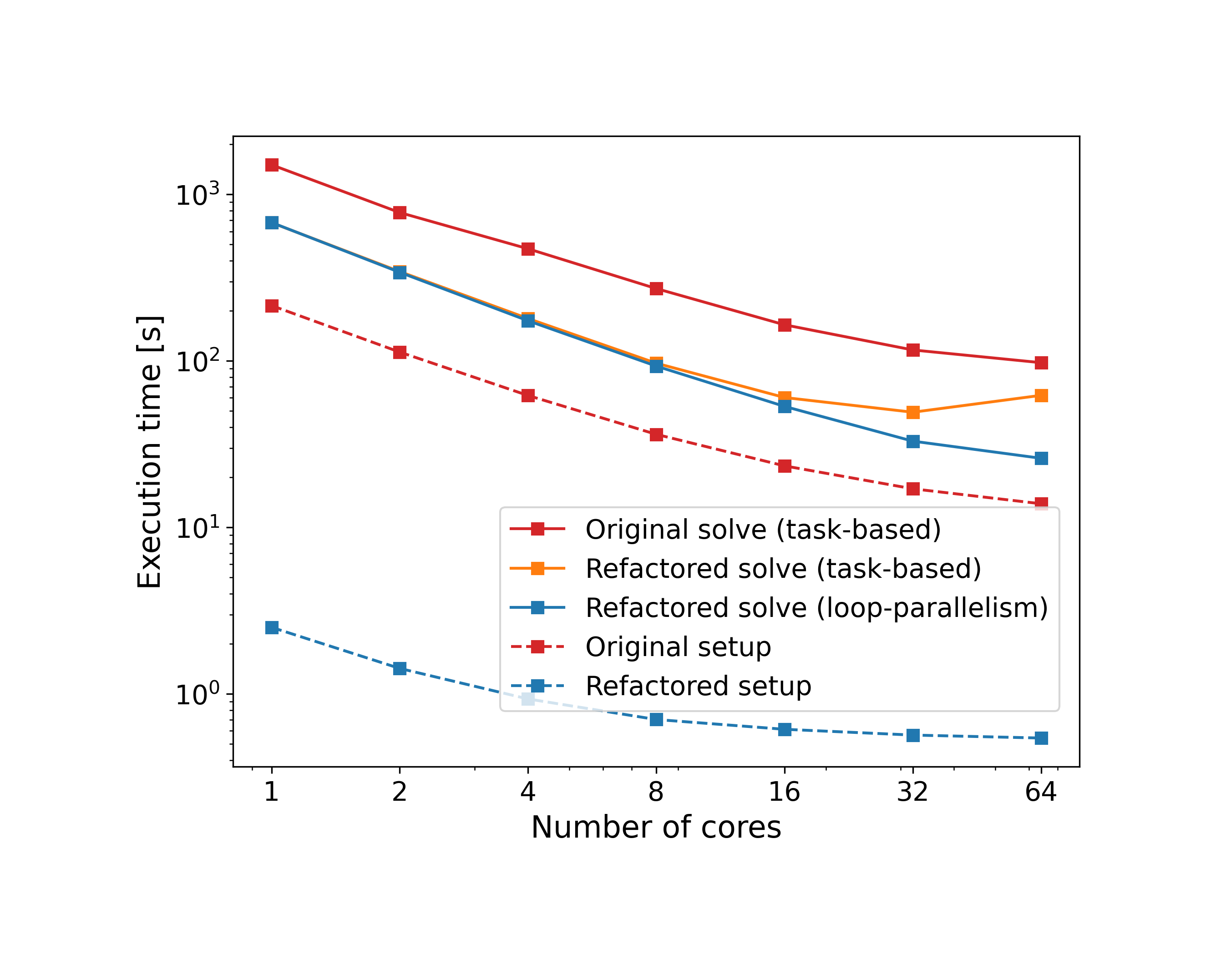}
			\end{minipage}
			\begin{minipage}[t]{0.48\textwidth}
				\centering
				\includegraphics[trim={2cm 2cm 2cm 2cm},clip,width=\textwidth]{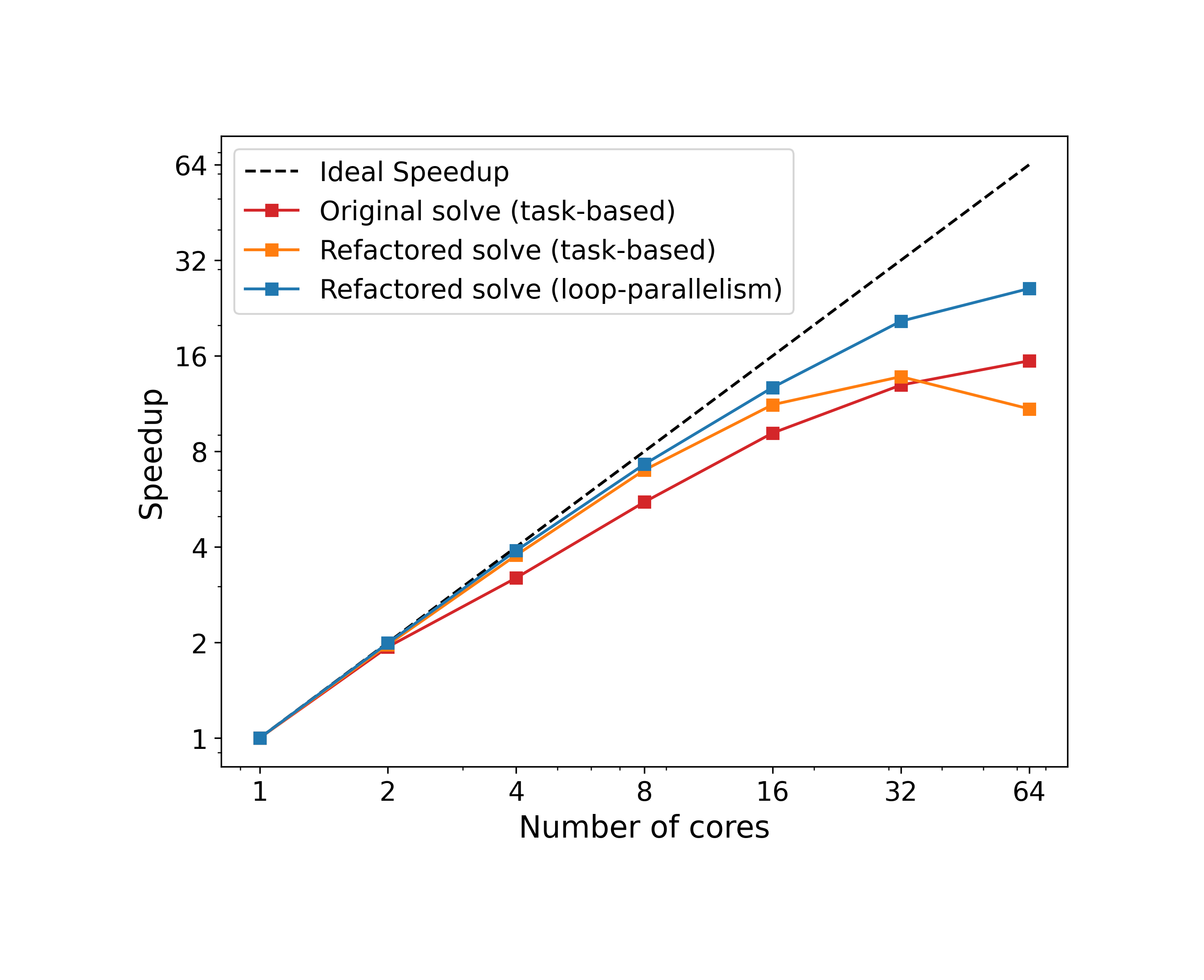}
			\end{minipage}
			\caption{\textbf{Strong scaling and runtime comparison of old and new implementation of GMGPolar.} Solver execution time in seconds (left) and speedup of solver times (right). \textbf{Czarny} geometry with \textbf{Give} stencil implementation and problem size $\mathbf{3073 \times 4096}$ with convergence criterion of a relative reduction of the initial residual by $1e8$.}
			\label{fig:strong_comparison_compute}
		\end{figure}
		
		Additionally, we compare old and new implementations on the Czarny geometry with a resolution of $3073 \times 4096$ nodes. 
		We consider the Give stencil implementation which was chosen for Version 1 in~\cite{LSK25}. In an intermediate step, we also compare the novel GMGPolar implementation with a task-based parallelism for the multigrid smoothers, which was implemented in Version 1. 
		From~\cref{fig:strong_comparison_compute}, we see first that both, setup and solve, timings could be substantially reduced. 
		For the setup phase we obtain a speedup of 86 and 26 for one and 64 cores, respectively. 
		For the solve phase we obtain speedups of 2.2 and 3.8 for one and 64 cores respectively. 
		We can furthermore state that the task-based parallelism reaches its limit of optimal performance with 16 to 32 cores and that the loop-based parallelism is better suited for the structured parallelism that we exploit in GMGPolar. 
		With 66~\% efficiency from one to 32 cores, the novel version scales very well for this test case.
		
		\subsection{Multigrid cycles, smoothing steps, and full multigrid initialization}\label{sec:cycles}
		\begin{table}[t!]
			\small
			\begin{center}
				\begin{tabular}[c]{cccccc}
					\toprule
					Initialization & Initial cycle & Cycle & its & Time (Init) & Time (MG) \\
					\midrule
					\bottomrule
					\multirow{6}{*}{No FMG} & - & \textbf{V(1,1)} &  \textbf{48}  & - & \textbf{77.64 s} \\
					& - & W(1,1) &  38  & - & 117.44 s \\
					& - & F(1,1) &  38  & - & 79.78 s \\
					& - & V(2,2) &  39 & - & 89.97 s \\
					& - & W(2,2) &  28  & - & 138.30 s\\
					& - & F(2,2) &  28  & - &  83.63 s\\
					\midrule
					\multirow{9}{*}{FMG} & 1x V & V(1,1) & 31 & 1.50 s & 36.21 s \\
					& 1x W & V(1,1) & 21 & 4.24 s & 33.81 s \\
					& 1x F & V(1,1) & 21 & 2.59 s & 33.99 s \\
					& 2x V & V(1,1) & 18 & 3.50 s & 28.12 s \\
					& 2x W & V(1,1) & 13 & 8.39 s & 21.02 s \\
					& \textbf{2x F} & \textbf{V(1,1)} & \textbf{13} &\textbf{5.01 s} & \textbf{21.00 s} \\
					& 3x V & V(1,1) & 15 & 5.29 s & 24.44 s \\
					& 3x W & V(1,1) & 12 & 12.62 s & 19.31 s \\
					& 3x F & V(1,1) & 12 & 7.48 s & 19.53 s \\
					\bottomrule
				\end{tabular}
				\caption{\textbf{Performance of different multigrid settings with and without full multigrid (FMG).} 
					\textbf{Czarny} geometry with \textbf{Take} stencil implementation and problem size $\mathbf{6145 \times 8192}$. The \textit{Initial cycle} column provides the cycle used with FMG initialization with the corresponding number of iterations. 
					The \textit{Cycle} column provides the cycle used in the multigrid iteration with the number of pre- and post-smoothing iterations. 
					The \textit{its} column provides the number of iterations until convergence of the multigrid scheme. 
					The \textit{Time (Init)} column provides the time for the FMG initialization and the \textit{Time (MG)} column provides the time of the multigrid scheme with convergence checks for an absolute residual smaller than $1e-16$.}   
				\label{tab:FMG}
			\end{center}
		\end{table}
		
		After having considered scaling properties and computational performance in the previous sections, we now consider the novel algorithmic features available with GMGPolar v2. 
		Therefore, we compare multigrid $V-$cycles with newly implemented $W$- and $F$-cycles. 
		Furthermore, we compare methods initialized by zero with an FMG initialization with either 1, 2, or 3 cycles of type $V$, $W$, or $F$. 
		As the Take implementation of the stencil performed best with respect to compute time, with reasonable increase in memory needs, we consider it here. 
		As a test case, we consider the Czarny geometry with a resolution of $6145 \times 8192$ nodes. 
		For a fair comparison with and without FMG, we require the iteration to reach an absolute residual of $1e-16$. 
		
		While we see from~\cref{tab:FMG} that without FMG, the $V$-cycle with just one pre- and postsmoothing step performs best with respect to the solve time, we can speed up the algorithm by a factor of three when using FMG with two initial $F$-cycles.
		
		Eventually, in~\cref{tab:finaltabspeed}, we provide the speedup for the Czarny geometry with a smaller grid size of approximately 780k nodes, as also considered in~\cite{BLK23}. 
		With the newly refactored GMGPolar, also using new multigrid features such as optimized initialization through FMG, we obtain substantial speedups compared to the old version. 
		With the Give approach, the speedup ranges between four and seven (for one to 16 cores) and with the Take approach between 16 and 18 (for one to 16 cores). 
		While the Take approach uses approximately the same amount of memory as the Give implementation in Version 1, the Give implementation of version 2 reduces the memory by approximately one third (36~\%).
		
		\begin{table}[H]
			\small
			\begin{center}
				\begin{tabular}[c]{cccc}
					\toprule
					\textnormal{Cores} &  1 & 4 & 16  \\
					\textnormal{Version} & &  &   \\            
					\midrule
					\bottomrule
					v1, Give & 83.96 s & 27.81 s & 13.79 s \\
					v2, Give, FMG (2x F) & 20.21 s (\textbf{4.15}) & 5.61 s (\textbf{4.96}) & 1.95 s (\textbf{7.07}) \\ 
					v2, Take, FMG (2x F) & 4.83 s (\textbf{17.38}) & 1.64 s (\textbf{16.96}) & 0.88 s (\textbf{15.67}) \\ 
					\bottomrule            
				\end{tabular}
				\caption{\textbf{Speedup of the novel GMGPolar.} 
					\textbf{Czarny} geometry with \textbf{Take} and \textbf{Give} stencil implementation and problem size $\mathbf{769 \times 1\,024}$. Solver timings shown for 1, 4, and 16 cores for the novel GMGPolar (v2) compared to the prior version (v1) in seconds and speedup shown between parentheses in bold face.}
				\label{tab:finaltabspeed}
			\end{center}
		\end{table}

		\subsection{GMGPolar in preconditioned conjugate gradients}\label{sec:pcg}
		
		\MK{In our prior publications as well as in the previous sections, GMGPolar was used as a standalone multigrid solver. However, in order to speed up convergence through an optimized construction of iterates, we can also use Krylov subspace methods such as the conjugate gradient (CG) method. In this section, we present preliminary and experimental results using GMGPolar as a preconditioner for CG (PCG-GMGPolar). In this setting, we consider the system
			\begin{align}\label{eq:pcgsystem}
				M^{-1}A_{ex}u=M^{-1}f_{ex},
			\end{align}
			where $A_{ex}$ and $f_{ex}$ are the extrapolated system and right hand side, as described briefly above and in more detail around~\cite[Eq.~(4.10)]{KKR21} and~\cite[Eq.~(32)]{LSK25}.
			
			In PCG, we first compute $r_0=f_{ex}-A_{ex}u_0$ and then either apply $M^{-1}$ or solve $Mz_0=r_0$ to obtain the preconditioned residual, which is used to initialize the search direction for solving~\eqref{eq:pcgsystem}. In our experiments, we use the nonextrapolated system matrix $M=A$ as the preconditioner, as it resulted in a shorter time-to-solution than $M=A_{ex}$. Within each PCG iteration, an approximate solution to $M z_k = r_k$ is obtained by performing a single FMG 1xF-cycle iteration.
			This provides a computationally efficient yet sufficiently accurate solution to the preconditioning step.
			
			In the following, we present results for standalone GMGPolar and PCG-GMGPolar using an initial approximation that is either set to zero or obtained via 1, 2, or 3 FMG iterations prior to the start of the multigrid or CG iterations.
			In~\cref{fig:pcg_gmgpolar}, we observe that the algorithm can achieve computational speedups of approximately 1.8 to 2.5, particularly when the initial FMG iterations alone are insufficient to reach the required tolerance.
			
			\begin{figure}[h!]
				\centering
				\includegraphics[trim={0cm 0cm 0cm 0cm},clip,width=0.6\textwidth]{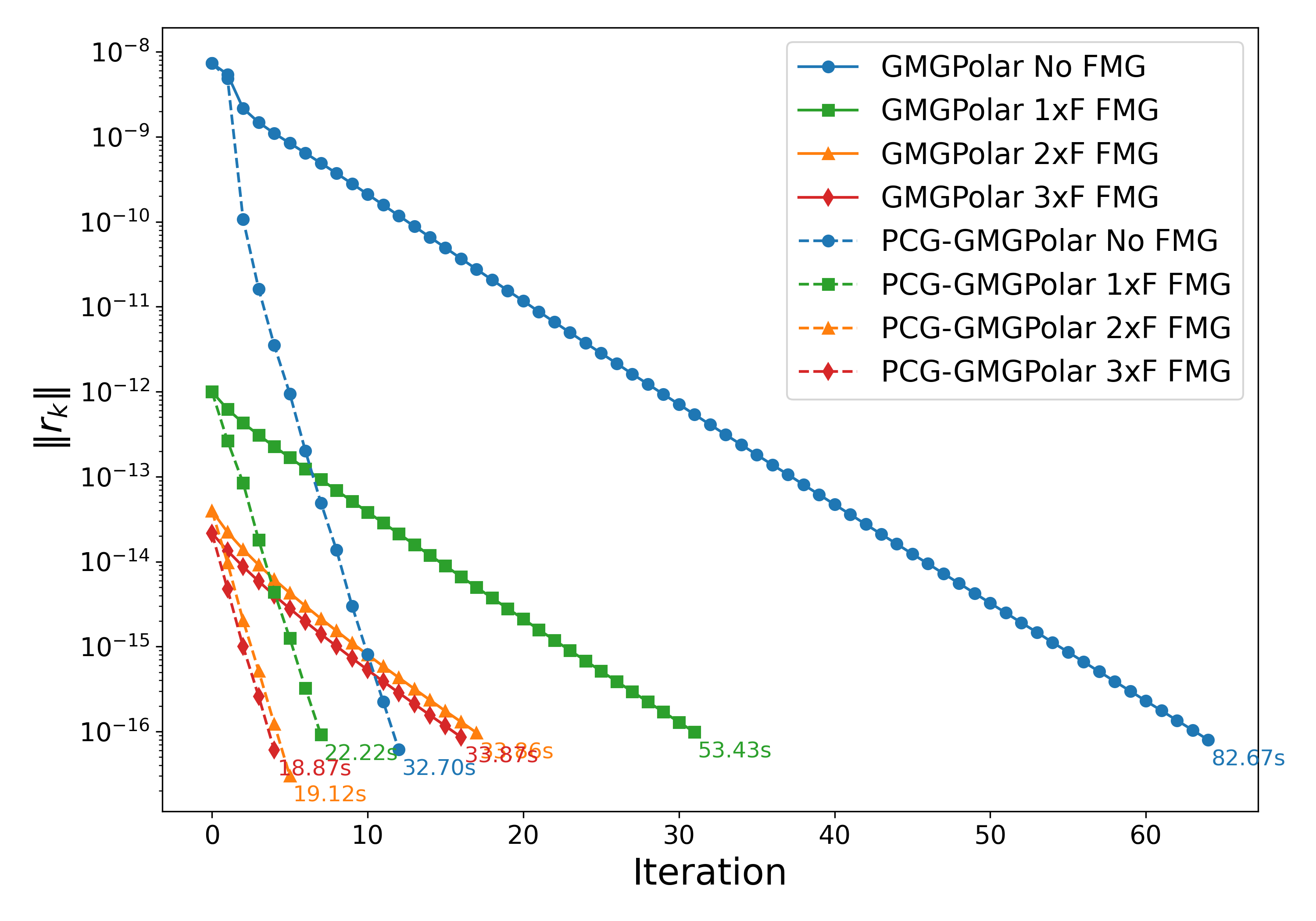}
				\caption{\MK{\textbf{Comparison of the experimental branch of standalone GMGPolar and GMGPolar in PCG.} The plot shows the convergence behavior in terms of the norm of the residual and the number of PCG iterations. Solver execution time in seconds shown after the last iteration. \textbf{Shafranov} geometry with \textbf{Take} stencil implementation and problem size $\mathbf{6145 \times 8192}$ with convergence criterion of an absolute residual smaller than $1e-16$.}}
				\label{fig:pcg_gmgpolar}
			\end{figure}
			
			In contrast to the prior results which have been fully merged to the productive main branch, using GMGPolar with PCG is still experimental and available on development branch \texttt{v2\_paper\_conjugate\_gradient}\footnote{\url{https://github.com/SciCompMod/GMGPolar/tree/paper_v2_conjugate_gradient}}. In between production version 2.0 and this branch some other minor adaptations and optimizations have taken place. A direct comparison with the prior release version has thus to be conducted with care. In~\cref{tab:finaltabspeed_pcg}, we have thus recomputed the row of the Take approach from~\cref{tab:finaltabspeed} and additionally evaluated the PCG version of GMGPolar. First of all, we observe some minor speedups from approximately 16-17 before to 17-20 with the minor improvements. However, we also see that the additional speedup through PCG yields an overall speedup of more than 37 in serial execution and more than 25 in 16-core execution.

			\begin{table}[H]
				\small
				\begin{center}
					\begin{tabular}[c]{cccc}
						\toprule
						\textnormal{Cores} &  1 & 4 & 16  \\
						\textnormal{Version} & &  &   \\          
						\midrule
						\bottomrule
						v2.2-experimental, Take, FMG (2x F) & 4.08 s (\textbf{20.57}) & 1.43 s (\textbf{19.45}) & 0.78 s (\textbf{17.68}) \\ 
						\bottomrule            
						v2.2-experimental-PCG, Take, FMG (2x F) & 2.26 s (\textbf{37.15}) & 0.88 s (\textbf{31.60}) & 0.55 s (\textbf{25.07}) \\ 
						\bottomrule            
					\end{tabular}
					\caption{\MK{\textbf{Speedup of the experimental branch of standalone GMGPolar and GMGPolar in PCG.} 
							\textbf{Czarny} geometry with \textbf{Take} stencil implementation and problem size $\mathbf{769 \times 1\,024}$. Solver timings shown for 1, 4, and 16 cores compared to the prior version (v1) in seconds and speedup shown between parentheses in bold face; cf.~\cref{tab:finaltabspeed}.}}
					\label{tab:finaltabspeed_pcg}
				\end{center}
			\end{table}
		}
		
		\section{Conclusion}\label{sec:conc}
		
		In this paper, we presented a completely refactored version of GMGPolar, an optimized state-of-the-art matrix-free multigrid solver that has been designed for complex 2D cross sections of tokamaks. GMGPolar has been developed to minimize the memory footprint, to allow fast and scalable execution for curvilinear coordinate representations and to achieve higher order approximations on tensor product structured grids.
		
		With the improved stencil Give implementation, GMGPolar minimizes the already small memory requirements by reducing it by approximately 36~\%. \MK{This yields an 8- to 14-fold reduction of memory compared to the spline solver, as demonstrated in~\cite{BLK23}, efficiently allowing to compute many different cross sections on a single compute node.}
		With improved weak and strong scaling, both stencil implementations, Take and Give, realize substantial speedups compared to the prior implementation. 
		For a use case as considered in~\cite{BLK23}, the Give implementation attains speedups between four and seven while the Take implementation attains speedups of approximately~16~to~18. \MK{In an experimental setting, we additionally considered GMGPolar as a preconditioner in the conjugate gradient method, which yielded an additional speedup factor of 1.8 to 2.5. In this experimental setup, we obtained speedups of more than 37 in serial execution and of more than 25 when executing on 16 cores.}
		
		While in~\cite{BLK23}, GMGPolar was found to represent "a compromise between relatively fast execution and high order of approximation", when compared to other state-of-the-art solvers, the novel version of GMGPolar combines an even reduced memory footprint with faster execution and better scalability.
		It can directly be used for domains without X-points, such as described in~\cref{fig:sol_polar6}, or in a multipatch decomposition for a fast and precise computation on the core part of the domain as presented in~\cite{vidal2025local}. 
		In addition, GMGPolar's object-oriented redesign offers a more intuitive use of the taylored geometric multigrid for physicists and plasma fusion engineers. 
		Future research will include porting the application to GPU accelerators and considering domains with X-points.
		
		\section*{Acknowledgements \& Funding}
		
		The authors gratefully acknowledge the scientific support and HPC resources provided by the German Aerospace Center (DLR). 
		The HPC system CARA is partially funded by "Saxon State Ministry for Economic Affairs, Labour and Transport" and "Federal Ministry for Economic Affairs and Climate Action".\\
		
		This project has received funding from the European High Performance Computing Joint Undertaking under grant agreement n°101144014.\\
		
		Funded by the European Union. Views and opinions expressed are however those of the author(s) only and do not necessarily reflect those of the European Union or the EuroHPC JU. Neither the European Union nor the granting authority can be held responsible for them.
		
		\begin{figure}[h!]
			\centering
			\includegraphics[width=1.0\textwidth]{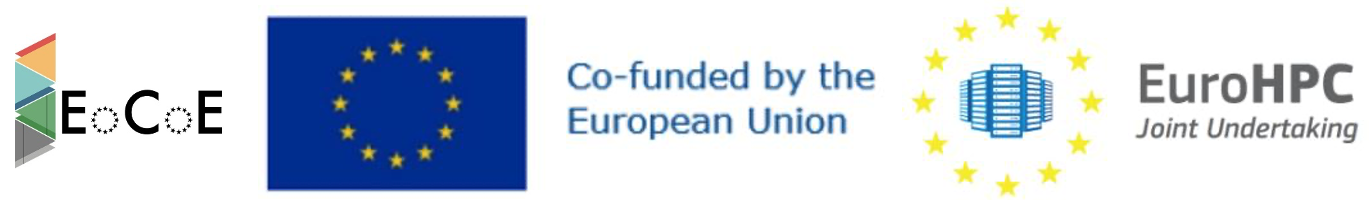}
		\end{figure}

		\section*{Author Contributions}
		
		\noindent\textbf{Conceptualization:} Julian Litz, Martin Kühn\\
		\noindent\textbf{Data Curation:} Julian Litz\\
		\noindent\textbf{Formal Analysis:} Julian Litz, Martin Kühn\\
		\noindent\textbf{Funding Acquisition:} Martin Kühn\\
		\noindent\textbf{Investigation:} All authors\\
		\noindent\textbf{Methodology:} All authors\\
		\noindent\textbf{Project Administration:} Martin Kühn\\
		\noindent\textbf{Resources:} Martin Kühn\\
		\noindent\textbf{Software:} Julian Litz\\
		\noindent\textbf{Supervision:} Joscha Gedicke, Martin Kühn \\
		\noindent\textbf{Validation:} All authors \\
		\noindent\textbf{Visualization:} Julian Litz, Martin Kühn\\
		\noindent\textbf{Writing – Original Draft:} Julian Litz, Martin Kühn\\
		\noindent\textbf{Writing – Review \& Editing:} All authors
		
		\bibliographystyle{elsarticle-num}

		\newgeometry{left=2cm,top=2cm,bottom=2cm}
		\pagestyle{empty}

		\section*{Appendix}
		For potential users, we provide the most important simulation parameters in~\cref{tab:parameters}.
		\begin{table}[h!]
			\centering
			\renewcommand{\arraystretch}{1.2}
			\small
			\begin{tabular}{|c|l|p{10cm}|}
				\hline
				\multicolumn{1}{|c|}{\textbf{Type}} & \textbf{Parameter} & \textbf{Description} \\
				\hline
				\parbox[t]{3mm}{\multirow{8}{*}{\rotatebox[origin=c]{90}{\large General}}} 
				& \ttvar{verbose} & Controls output verbosity. \\
				& \ttvar{paraview} & Enables Paraview output files. \\
				& \ttvar{maxOpenMPThreads} & Maximum OpenMP threads. \\
				& \ttvar{stencilDistributionMethod} & Stencil distribution: `Take' or `Give'. \\
				& \ttvar{cacheProfileCoefficients} & Caches profile coefficients $\alpha$ and $\beta$. \\
				& \ttvar{cacheDomainGeometry} & Caches transformation coefficients $a^{rr}$, $a^{r \vartheta}$ and  $a^{\vartheta \vartheta}$. \\
				& \ttvar{DirBC_Interior} & Interior boundary condition: Across-the-origin or Dirichlet. \\
				\cline{1-3}
				\parbox[t]{3mm}{\multirow{7}{*}{\rotatebox[origin=c]{90}{\large Polar Grid}}} 
				& \ttvar{R0} & Generalized radius of the innermost circle. \\
				& \ttvar{Rmax} & Generalized radius of the outermost circle. \\
				& \ttvar{nr_exp} & Number of discretization points in radial dimension. \\
				& \ttvar{ntheta_exp} & Number of discretization points in angular dimension. \\
				& \ttvar{anisotropic_factor} & Anisotropic refinement radius. \\
				& \ttvar{divideBy2} & Refines grid globally \ttvar{divideBy2} times to obtain identical grids for scaling experiments. \\
				\cline{1-3}
				\parbox[t]{3mm}{\multirow{13}{*}{\rotatebox[origin=c]{90}{\large Multigrid Settings}}} 
				& \ttvar{FMG} & Enables full multigrid / nested iteration for initial approximation. \\
				& \ttvar{FMG_iterations} & Number of FMG iterations. \\
				& \ttvar{FMG_cycle} & FMG Cycle type: $V$-, $W$-, or $F$-cycle. \\
				& \ttvar{extrapolation} & Extrapolation: None, implicit or full grid smoothing. \\
				& \ttvar{maxLevels} & Maximum multigrid levels. \\
				& \ttvar{preSmoothingSteps} & Pre-smoothing steps. \\
				& \ttvar{postSmoothingSteps} & Post-smoothing steps. \\
				& \ttvar{multigridCycle} & Multigrid Cycle type:$V$-, $W$-, or $F$-cycle. \\
				& \ttvar{residualNormType} & Residual norm type: $\|\cdot\|_2$, weighted $\|\cdot\|_{l_2}$, or $\|\cdot\|_{\infty}$. \\
				& \ttvar{maxIterations} & Maximum multigrid iterations. \\
				& \ttvar{absoluteTolerance} & Absolute tolerance for convergence. \\
				& \ttvar{relativeTolerance} & Relative tolerance for convergence. \\
				\cline{1-3}
				\parbox[t]{3mm}{\multirow{7}{*}{\rotatebox[origin=c]{90}{\large Test Problem}}}
				& \ttvar{geometry} & Cross section shape: Shafranov, Czarny, Culham, etc. \\
				& \ttvar{alpha_jump} & Radius of rapid decay for density profile. \\
				& \ttvar{kappa_eps} & Geometry elongation. \\
				& \ttvar{delta_e} & Outward radial displacement of flux center. \\
				& \ttvar{problem} & Defines the solution: Cartesian, Polar, Multi-scale. \\
				& \ttvar{alpha_coeff} & Alpha coefficient: Poisson, Sonnendrucker, Zoni. \\
				& \ttvar{beta_coeff} & Beta coefficient: Zero or inverse of \ttvar{alpha_coeff}. \\
				\hline
			\end{tabular}
			\caption{\textbf{Summary of simulation parameters with descriptions.}}
			\label{tab:parameters}
		\end{table}

		\restoregeometry
		
	\end{document}